\documentclass{article}
\usepackage{amsmath,amsfonts,amssymb}
\usepackage{authblk}
\usepackage{graphicx}
\usepackage[a4paper, margin=1in]{geometry}
\usepackage{hyperref}
\usepackage{csquotes}
\DeclareUnicodeCharacter{2009}{\,}
\title{Generation and Enhancement of Persistent Nanoscale Magnetization in All-Dielectric Metasurfaces by Optically Injected and Localized Free Carriers}
\author[*,1]{Shivaksh Rawat}

\author[1]{Samyobrata Mukherjee}

\author[*,1]{Gennady Shvets}

\affil[1]{School of Applied and Engineering Physics, Cornell University, Ithaca, NY 14850, USA}
\affil[*]{Email: sr939@cornell.edu, gshvets@cornell.edu}

\date{}
\begin{document}
\maketitle
\begin{abstract}
Time-varying dielectric metasurfaces that support sharp optical resonances with nontrivial electromagnetic field distributions constitute a unique platform for realizing temporal interfaces for metasurface-guided waves (MGWs). Rapidly changing metasurface resonance enables frequency conversion and temporal scattering of a concurrently propagating MGW. Using analytical methods and electromagnetic simulations, free carriers are generated locally to create frequency-shifted infrared MGWs. Such time interfaces can be utilized to generate large, highly localized quasistatic magnetic fields within the metasurfaces. The resulting nanoscale magnetization, supported by the residual circulating currents, persists for several optical cycles after the departure of the time-scattered MGWs. During the rectification process, the initial electromagnetic energy of the injected MGWs is partitioned between the temporally scattered MGWs, the residual motion of the free carriers, and a quasistatic magnetic field.
\end{abstract}

\section{Introduction}
Time-varying media have been widely studied in plasma physics for several decades. The theory of frequency upshifting \cite{Eli1974, Dawson1989PRL} and spectral broadening \cite{Eli1973} of an electromagnetic pulse was originally developed for gaseous plasmas. Rapid free-carrier (FC) generation by a laser pulse via photoionization upshifts the pulse's frequency \cite{Downer1991, Savage1992}. Recent research has focused on time-varying optical media, offering a new platform for studying phenomena such as photon acceleration (PA) \cite{Maxim2019NatComm, Boyd2021PA_Nano_lett}, wave amplification in momentum bandgaps using photonic time crystals \cite{Tretyakov2023SciencePTC, Asadchy2024NatPhoto}, negative wave extinction \cite{Maxim2019Optica}, phase conjugation and negative refraction \cite{Cummer2010PRL}, and surface wave control using time interfaces (TIs) \cite{Tretyakov2023Nanophotonics, Alu2020PRL}, among several other explorations of novel physical phenomena \cite{Engheta2020Optica, Inv_prism2018OptLett, Tirole2023NatPhys, Moussa2022NatPhys, Alu_Photo_coll_2022NatPhys}. We refer the reader to the review article in Ref. \cite{Pendry2022ReviewPTC} for a comprehensive overview of time-varying photonics.

One of the many interesting applications of time-varying photonics is on-demand up- and down-conversion of the incident frequency, with potential uses in optical signal processing and communications. Lately, the large optical near-field amplification observed in doped transparent conducting oxides (TCOs) has been used to enhance light-matter interactions \cite{Alam2016Science, Boyd2018NatPhoto}. Intense pump pulse incidence on a TCO leads to temperature modulation of the electrons in the conduction band, which in turn affects the plasma frequency $\omega_\mathrm{p}$, and has allowed for tunable redshifting of the probe pulse \cite{Boyd2021PA_Nano_lett, Brener2019NatPhys}. Dynamic tuning of the probe frequency with selective blue- or red-shifting has been observed with two-color (UV and IR) excitation in an AZO thin film \cite{Shalaev_Two_color2017NatComms}. However, the need for two pump pulses at different frequencies renders this frequency-shifting scheme complex and difficult to implement. Moreover, these approaches rely on coupling the probe wavelength to the epsilon-near-zero (ENZ) mode of the substrate, constraining the accessible spectral range as a function of doping and geometric parameters, which cannot be tuned dynamically during operation. Furthermore, a significant drawback of TCOs is their extreme lossy nature in the ENZ regime, where they are typically operated \cite{Alam2016Science}.

We propose tuning the resonance frequency of a metasurface to realize a time interface in a structured optical medium. Backed by numerical simulations and analytic perturbation theory, we demonstrate tunable red- and blueshifting of the metasurface resonance using localized free carrier generation by a pump laser beam in a hot spot at each meta-atom. We employ this resonance-shifting mechanism to create a TI for a mid-IR (MIR) metasurface-guided wave (MGW). The MGW is a transversely confined metasurface mode that propagates along the metasurface and does not leak into the air above or the dielectric substrate below. Local FC generation by a near-IR (NIR) pump pulse in the meta-atom, on a timescale comparable to the optical cycle of the MGW, shifts the metasurface resonance and creates a sharp TI.  Furthermore, we investigate the energy dynamics of the propagating MGW across the TI. We find that after the TI, the electromagnetic energy of the launched MGW is divided between the electromagnetic energy of the temporally scattered MGW, the kinetic energy of the free carriers, and the magnetic energy in a quasistatic magnetic field created in the hot spot by the TI. The FCs generated in the hot spot are accelerated by the electromagnetic fields of the MGW during the TI, leading to the formation of quasistatic currents that circulate within the hot spot and persist after the TI. Consequently, the resonantly enhanced time-varying AC magnetic field of the MGW is efficiently rectified, yielding a quasistatic magnetic field localized in the hot spot. Remarkably, even when the effect of electron scattering losses is considered, this rectified quasistatic magnetic field persists for several scattering times after the TI.

Nonlinear optical properties of certain materials have been used to generate (i) quasi-static electric fields from time-varying fields via electro-optic rectification \cite{Bosshard1995, Fulop2020}, and (ii) light-induced magnetization via the Inverse Faraday Effect (IFE) \cite{IFE1965PRL, IFE2024APR}. However, the typically weak nonlinear optical properties of common optical materials constrain their applications. Besides, the IFE typically generates instantaneous magnetic fields proportional to the pump pulse intensity. Recently, optical pump-induced ferromagnetic order has also been reported in two-dimensional materials \cite{Hao2022}, and laser-assisted nanostructuring of magnetic materials has been used to generate structured magnetic fields \cite{Levati2025}. Here, we propose an efficient method for rectifying propagating AC magnetic fields using a time interface created by localized FC generation in an all-dielectric metasurface, which is agnostic to the nonlinear optical properties of the constituent materials. The rectification of AC magnetic fields from a propagating electromagnetic wave has been studied in the context of a uniform plasma \cite{Mori_1988_B_Field}. However, to the best of our knowledge, this is the first proposal of the use of a TI in the rectification of an AC magnetic field to generate a persistent quasistatic magnetic field that leverages the resonant enhancement provided by a nanostructured metasurface and is dependent on the three-dimensional optical fields of the MGW and the hot spot. Nanostructured three-dimensional magnetic fields have potential applications in spintronics and computing \cite{Fernandez2017}.

The rest of the manuscript is organized as follows. An all-dielectric metasurface (semiconductor meta-atoms atop an infrared transparent substrate) that will be used throughout this work as a platform to demonstrate a TI is described in Section~\ref{sec:MS_resonances}. A specific metasurface design supporting high-quality factor MIR resonances and highly localized \enquote{hot spots} of the electric field intensity that can be used for localized free carrier generation (LFCG) is described in Sect.~\ref{subsec:design}. Several perturbative techniques for deriving rigorous analytic expressions for the resonance-frequency shifts due to local FC generation are described in Sect.~\ref{subsec:freq_shift}, which account for the vector nature of electromagnetic fields within meta-atoms. We demonstrate that, by controlling the density and spatial distribution of the generated carriers, either red- or blue-shifting of a metasurface resonance can be achieved, with specific examples for a germanium-based metasurface presented in Sect.~\ref{subsec:Ge_example}. In Sect.~\ref{sec:IIIA}, we detail our approach to the modeling of laser-driven free carrier generation in a dielectric. The effects of FC generation on a dispersive Drude-Lorentz medium are described in Sect.~\ref{subsec:TVDLM}, where we rigorously derive the expressions for the currents due to the newly created FCs. In Sect.~\ref{subsec:Energy} we derive an analytic expression for the total energy density in a time-varying Drude-Lorentz medium, where we find that the total energy in the system remains unaffected by the FCs, but rather a re-partitioning of the energy between the EM fields and the free carriers takes place. In Sect.~\ref{subsec:IV_A}, we demonstrate the effects of a sharp TI on a metasurface-guided MIR mode using the pump-probe approach via time-domain simulations. An ultrashort high-frequency pump pulse can produce temporal reflections via localized metallization for a lower-frequency MIR mode while also redshifting the mode. We study the energy dynamics of the MGW mode across a time interface in Sect.~\ref{subsec:IV_b} and describe the partitioning of the MGW's energy after the TI. In Sect.~\ref{subsec:IV_c} we describe a quasistatic magnetic field that is generated as a result of the TI via rectification of the AC magnetic fields of the propagating MGW. Our findings are summarized in Section~\ref{sec:conclusions}.

\section{Controlling MS Optical Properties via FC Generation}\label{sec:MS_resonances}
The most intuitive and technically straightforward approach to rapidly modulating the refractive index of a solid material is to use a laser beam to generate high carrier densities. The process of free carrier generation can be further enhanced by using resonant dielectric or semiconductor metasurfaces. Nonlinear light-matter interactions inside such metasurfaces are significantly enhanced by the combination of the local field enhancement at the \enquote{hot spots} with a high-quality factor ($Q$) resonance that traps light for long time periods \cite{Maxim2019NatComm, Maxim2023NatComms}. Moreover, dielectric metasurfaces have significantly higher optical damage thresholds compared to their plasmonic counterparts, which suffer higher ohmic losses \cite{Wu_Shvets_2014NatComm, Yuri2018NanoLett} and have higher local field enhancement due to the plasmonic resonance, which would, in turn, render them more susceptible to damage. Thus, an all-dielectric high-$Q$ metasurface with a tunable resonance via localized free-electron generation by an intense pump pulse and the probe experiences the resulting change of the metasurface properties \cite{Ge_Opt2019}. In the following sections, we introduce a metasurface-based approach that uses spatially localized free-carrier generation to tune the metasurface resonance frequency.

\subsection{Metasurface Design}
\label{subsec:design}
\begin{figure}[t]
 \centering
\includegraphics[width=90mm]{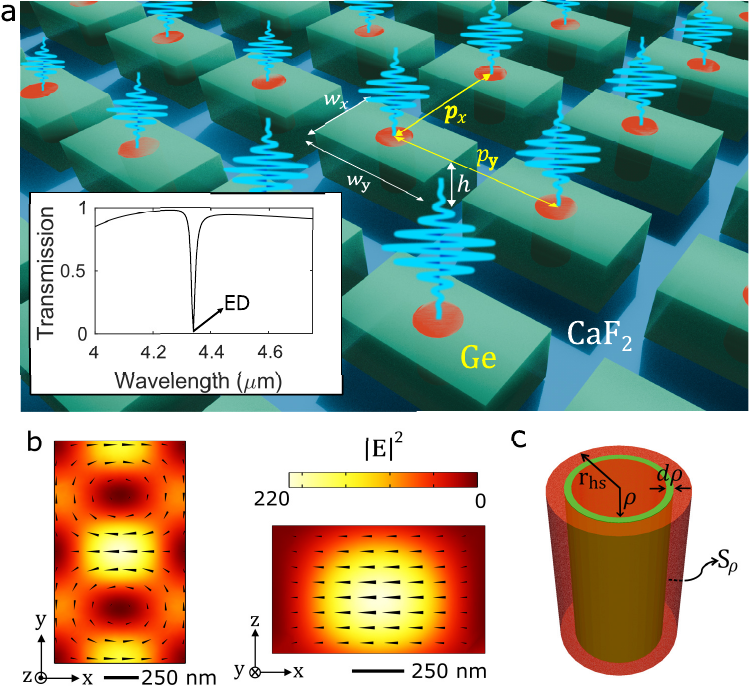}
 \caption{\textbf{a.} Schematic of a metasurface comprising rectangular semiconductor blocks periodically arranged on an infrared-transparent substrate. Red spots: cylindrical region where a pump pulse generates free-carriers. Inset: metasurface transmission spectra for normally incident $x$-polarized mid-infrared light. \textbf{b.} Electric field (arrows) and its intensity $|\mathbf{E}|^2$ (color-coded) distribution at the electric dipolar (ED) resonance in the $x$-$y$ (left panel) and $x$-$z$ (right panel) planes drawn through the middle of a meta-atom. \textbf{c.} Schematic for the perturbative calculation of the resonance frequency shift produced by free-carrier generation inside the hot-spot cylinder with radius $\mathrm{r_{hs}}$ using annular rings (shown in green). Materials: Germanium (Ge, $\mathrm{n_{Ge} \approx 3.98}$ \cite{Amotchkina2020}) for meta-atoms, Calcium Fluoride (CaF$_2$, $\mathrm{n_{CaF_2} \approx 1.4}$ \cite{Malitson1963}) for the substrate. Geometric parameters: $p_x=1.35\ \mu \mathrm{m},\ p_y=2\ \mu \mathrm{m}$, $w_x \times w_y \times h$: $1\ \mu \mathrm{m} \times 1.7\ \mu \mathrm{m} \times 0.6\ \mu \mathrm{m}$, and $\mathrm{r_{hs}}=210\ \mathrm{nm}$.}
 \label{Fig1}
\end{figure}
The metasurface design comprises a two-dimensional array of high-index rectangular semiconductor blocks (the meta-atoms), unequally spaced by $p_{x(y)}$ in the $x(y)$ directions on an optically transparent substrate, as shown in Fig.~\ref{Fig1}. Note that even though the meta-atoms will be assumed to be made of the semiconductor material (e.g., germanium) for practical purposes, they will still be referred to as \enquote{all-dielectric metasurfaces} because of the nearly negligible frequency dependence of the refractive index of the constituent. We chose germanium because it offers near-zero loss and near-unity transmission in the mid-IR regime \cite{Amotchkina2020} and has an extremely high laser-induced damage threshold ($\sim$ 2.9 $\mathrm{TW/cm^2}$ for femtosecond pulses at $\sim2\ \mu \mathrm{m}$ \cite{LIDT_Ge_mid_IR}). The design parameters of the metasurface are detailed in the caption of Fig.~\ref{Fig1}. The refractive indices of the substrate and meta-atom material (CaF$_2$ and Ge, respectively) are assumed to be frequency independent for the range of mid-infrared (MIR) frequencies of the probe.

The optical properties of the metasurface are shown in the inset of Fig.~\ref{Fig1}a, where the numerically calculated transmission spectrum is plotted for mid-IR frequencies. The dip in the transmission spectrum corresponds to a high-$Q$ ($Q\sim 287$) electric dipolar resonance (marked ED at $\lambda_{\rm ED} \sim 4.34\ \mu \rm{m}$). The electric field (direction and intensity) for the ED resonance in the meta-atom's $x$-$z$ midplane is shown in Fig.~\ref{Fig1}b. The electric classification of the resonance is based on the vectorial nature of the electric field and the non-rotational nature of the electric field at $\lambda_{\rm ED}$ indicates an ED resonance. In addition to its narrow spectral width $\delta \omega \equiv \omega_0/Q$, the ED resonance has a second important characteristic: a strongly enhanced and highly localized electric field intensity, as shown in Fig.~\ref{Fig1}b. As will be explained in Section~\ref{subsec:freq_shift}, the amount of resonance frequency shift $\Delta \omega$ produced by the generation of free carriers is proportional to the fraction of the energy of the electric field contained in the hot spot. Similarly, the effect of this frequency shift on the probe light stored inside the metasurface (i.e., the efficiency $\eta$ of the TI) is proportional to the ratio of the resonance frequency shift $\Delta \omega$ and the resonance bandwidth, i.e., $\eta \sim Q \Delta \omega$. Thus, both the resonant field enhancement and the narrow bandwidth of the high-$Q$ ED resonance are beneficial to creating an efficient TI.

\subsection{Shifting the Resonance Frequency of Spectrally-selective Metasurfaces: Perturbation Theory}\label{subsec:freq_shift}
In this section, we develop a general formalism to calculate the resonance frequency shift of a high-$Q$ all-dielectric metasurface using LFCG. LFCG reduces the dielectric permittivity of the constitutive material from $\varepsilon_{\rm ini}$ to $\varepsilon_{\rm fin}\equiv \varepsilon_{\rm hs}$ in the hot spot. Two LFCG regimes will be considered: (i) the perturbative regime corresponding to a low density of the free carriers ($|\varepsilon_{\rm ini} - \varepsilon_{\rm fin}| < \varepsilon_{\rm ini}$), and (ii) the non-perturbative regime of local \enquote{metallization} corresponding to the high density of the free carriers ($\varepsilon_{\rm fin} \ll -\varepsilon_{\rm ini}$).

Small localized modifications in the permittivity of a material inside a resonator alter its resonance frequency, and this change in resonance frequency may be understood as a perturbative effect \cite{Waldron1960,bethe_schwinger_1943, Schwinger:1943,hauser_electromagnetism_1971}. Early work required knowledge of the exact perturbed electromagnetic fields (which are generally unknown) to predict the resonance shift. More recent work uses techniques borrowed from time-independent perturbation theory \cite{Griffiths2018} in quantum mechanics to predict resonance frequency shifts in perturbed electromagnetic cavities \cite{Joannopoulos1995, Johnson2002, Johnson2008, Hossein2013}. We utilize the formalism outlined in Ref. \cite{bethe_schwinger_1943} and derive an equation, which accounts for the full-vector electric field within the perturbed hot spot, and accurately predicts fractional changes in resonance frequency for small perturbations to the meta-atom. 

We consider a perturbation arising from a localized change in permittivity due to LFCG inside a hot spot located inside a semiconductor meta-atom from an initial relative permittivity $\varepsilon_{\mathrm{ini}}$ to a perturbed relative permittivity $\varepsilon_{\mathrm{fin}}(\mathbf{r})$, where $\varepsilon_{\mathrm{fin}}(|\mathbf{r}|>r_{\rm{hs}}) = \varepsilon_\infty$ and $\varepsilon_{\mathrm{fin}}(|\mathbf{r}|<r_{\rm{hs}}) = \varepsilon_{\mathrm{hs}}$ where $r_{\rm{hs}}$ is the radius of the cylindrical hot spot. The permittivity change inside the hot spot arises from changes in the free carrier density in the hot spot $\mathrm{N_e}(\mathbf{r},\mathrm{t})$, since $\varepsilon_\mathrm{hs}(\omega) = \varepsilon_{\infty} - \frac{\omega_\mathrm{p}^2}{\omega^2+\mathrm{i}\gamma\omega}$, and $\omega_\mathrm{p}^2=\frac{\mathrm{N_e e^2}}{\epsilon_0 \mathrm{m_e}}$. Here, $\varepsilon_{\infty}=\varepsilon_{\rm ini}\approx 16$ is the high-frequency response of bound electrons, $\epsilon_0$ is the permittivity of vacuum, $\mathrm{m_e}=0.041\ \mathrm
m_0$ is the effective mass of the free carriers near the $\Gamma$ point of the conduction band \cite{van2011book}, and an arbitrary electronic scattering rate, $\gamma^{-1}=100\ \mathrm{fs}$ is assumed in the hot spot to account for various losses. Following the standard perturbation theory~ \cite{bethe_schwinger_1943,hauser_electromagnetism_1971}, we derived the resonance frequency shift $\Delta \omega$ using only the unperturbed electromagnetic fields  (see Section 1 of the supplementary information for details)
\begin{equation}
    \frac{\mathrm{Re}(\Delta\omega)}{\mathrm{Re}(\omega_0)}=\frac{\int_0^\mathrm{r_{hs}} \left\{ \oint_{\mathrm{S}}\epsilon_0\left( [\mathrm{\varepsilon_{ini}}-\mathrm{\varepsilon_{fin}}] |\mathbf{E}_{0\parallel}(\rho)|^2 +\left[\frac{1}{\mathrm{\varepsilon_{fin}}}-\frac{1}{\mathrm{\varepsilon_{ini}}} \right]|\mathbf{D}_{0\perp}(\rho)|^2\right)d\mathrm{S}\right\} \,\mathrm{d\rho}}{\int_\mathrm{V} (\mathbf{D}_0\cdot\mathbf{E}_0^*+\mathbf{B}_0\cdot\mathbf{H}_0^*)\ \mathrm{dV}},
\label{eq:delta_omega_frac_4}
\end{equation}
where the perturbed cylindrical hot spot comprises annular shells with radii $0<\rho<\mathrm{r_{hs}}$, thicknesses $\mathrm{d}\rho$, and area $S_{\rho}$ (see Fig. \ref{Fig1}c). The $\perp$/$\parallel$ components of $\mathbf{E}_0$/$\mathrm{D}_0$ are defined with respect to the surface $\mathrm{'S'}$ of the infinitesimally thin annular cylinder, and $'\mathrm{V}'$ is the metasurface unit cell volume. Though derived for small perturbations, eq.~(\ref{eq:delta_omega_frac_4}) provides several key insights. First, it follows that a small reduction in the dielectric permittivity of the hot spot due to LFCG blue-shifts the resonance frequency: $\Delta \omega >0$ if $0 < \varepsilon_{\rm fin} < \varepsilon_{\rm ini}$. Second, we note that the blueshift $\Delta \omega$ is expected to be particularly large when $\varepsilon_{\rm fin} \approx 0$, i.e., when the hot spot material enters the epsilon-near-zero (ENZ) regime. Third, Eq.~(\ref{eq:delta_omega_frac_4}) indicates that the sign of $\Delta \omega$ changes and the resonance redshifts when the hot spot becomes plasmonic ($\varepsilon_{\rm fin} < 0$), including the regime of complete metallization ($\varepsilon_{\rm fin} \ll -\varepsilon_{\rm ini}$). The comparison between the predictions from Eq.~(\ref{eq:delta_omega_frac_4}) and those of other methods is shown in Section S1 of the supplementary information (SI).

Although the regime of complete metallization constitutes a strong change in the material permittivity inside the hot spot from that of the rest of the meta-atom, we may still predict the resonance frequency shift. In fact, the calculation of the resonance frequency shift of a perturbed microwave cavity by Slater~ \cite{Slater1946} assumed the introduction of a small volume $\Delta \mathrm{V}$ of a perfect electric conductor (PEC) into the resonator and made specific predictions about blue- or red-shifts of the metasurface resonance. A detailed study of the effects of PEC introduction or localized metallization in the meta-atom is presented in Section S2 of the SI. Throughout the manuscript, we treat a PEC and complete metallization of the material inside the hot spot as interchangeable terms.

\subsection{Example: Resonance Frequency Shifting in Ge Metasurface}\label{subsec:Ge_example}
Modification of the local permittivity in the Ge block provides a mechanism to tune the metasurface resonance frequency. LFCG can modify the permittivity at the hot spot (red cylinders in Ge blocks in Fig.~\ref{Fig1}a). Specifically, we consider a hot spot with radius $\mathrm{r_{hs}=210}$ nm, where the local FC density varies rapidly. The value of $\mathrm{r_{hs}}$ is chosen based on the spatial profile of the FCs generated via tunneling photoionization due to a pump pulse. Tunneling ionization is a non-linear process and therefore allows the creation of a hot spot narrower than the diffraction-limited beam spot for a near-IR pump (see section S5 of the SI for details). An ultra-short, high-intensity pump pulse may be converted into a spatial array of focused pulses using a diffractive beam splitter \cite{Katz2018}, which would generate hot spots in the meta atoms. Alternatively, the hot spots may be created by local-field enhancement using the mid-IR probe itself. Deep subwavelength spatial localization of FC generation has been achieved using mid-IR pulses in the tunneling regime (e.g., the radius of the hot spot $\mathrm{r_{hs}}<\lambda/40$) \cite{Maxim2023NatComms}. In the rest of the manuscript, we shall focus on the pump-probe method using a near-IR pump for LFCG and a mid-IR probe.

\begin{figure}[t]
\centering
    \includegraphics[width=100mm]{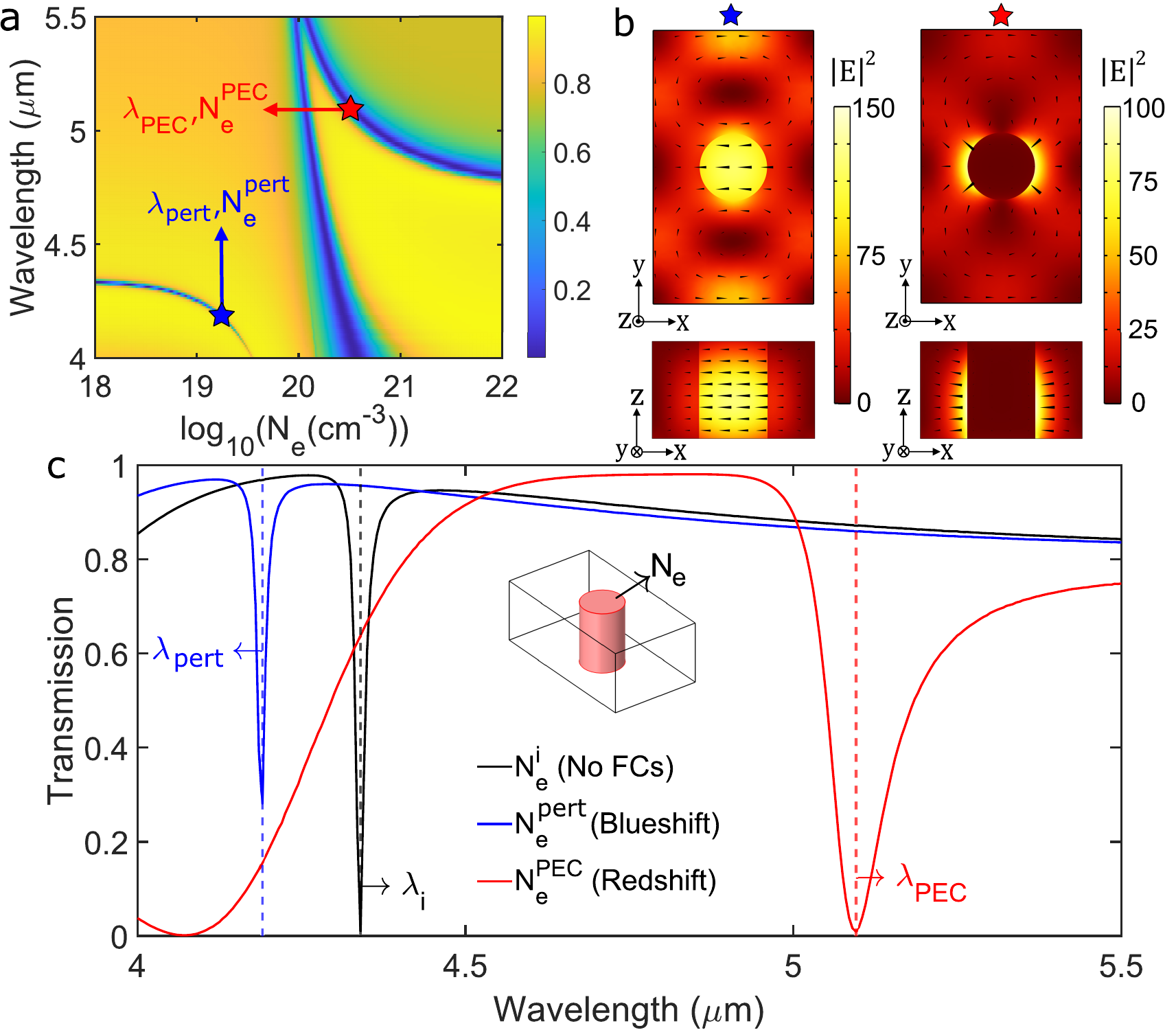}
    \caption{$\bf{a.}$ Metasurface transmission as a function of hot spot free carrier density ($\mathrm{N_e}(\mathbf{r},\mathrm{t})$) for normally incident x-polarized light. $\bf{b.}$ Intensity $\mathrm{|E|^2}$ enhancement (color), and $\mathbf{E}$ (black cones) of the ED resonance in the meta-atom's $x$-$y$ and $x$-$z$ mid-planes for $\lambda_{\mathrm{pert}}= 4.19\  \mu\mathrm{m}$, $\mathrm{N_e^{pert}}=1.6\times 10^{19}\  \mathrm{cm}^{-3}$ (blue star) and $\lambda_{\mathrm{PEC}}=5.09\  \mu\mathrm{m}$, $\mathrm{N_e^{PEC}}=3.2\times10^{20}\  \mathrm{cm}^{-3}$ (red star). The stars serve as guides to the eye. $\bf{c.}$ The black curve shows the metasurface transmission spectrum before FC generation ($\mathrm{N_e^{i}}=0\  \mathrm{cm}^{-3}$), while the blue (red) curve shows the spectrum when the carrier density is increased to $\mathrm{N_e^{pert}}$ ($\mathrm{N_e^{PEC}}$).}
    \label{Fig2}
\end{figure}
The evolution of the metasurface transmission spectrum (from COMSOL simulations) as the hot spot's FC density $\mathrm{N_e}(\mathbf{r},\mathrm{t})$ increases from $10^{18}\ \mathrm{cm^{-3}}$ to $10^{22}\ \mathrm{cm^{-3}}$ is shown in Fig.~\ref{Fig2}a. For $4\ \mu\mathrm{m}\leq\lambda\leq 5.5\ \mu\mathrm{m}$, the permittivity in the hot spot, $\varepsilon_\mathrm{hs}$, varies in the range $0<\varepsilon_\mathrm{fin}=\varepsilon_\mathrm{hs}<\varepsilon_\mathrm{ini}$ when $\mathrm{N_e} \leq 10^{20}\ \mathrm{cm^{-3}}$. We see in Fig.~\ref{Fig2}a that the dip in transmission corresponding to the metasurface ED resonance blue-shifts as $\mathrm{N_e}$ increases from $10^{18}$ to $10^{20}\ \mathrm{cm^{-3}}$ and $\varepsilon_\mathrm{fin} = \varepsilon_\mathrm{hs}$ decreases from $\varepsilon_\mathrm{ini}$ to $0$ (in agreement with the prediction of eq.\ref{eq:delta_omega_frac_4}). For $\mathrm{N_e}>10^{20}\ \mathrm{cm^{-3}}$, the ED resonance disappears and then reappears at much higher wavelengths, exhibiting a strong redshift. This is because for large FC density, the hot spot metallizes, strongly modifying the resonant fields. Even though such modifications are beyond the perturbative regime, eq.~\ref{eq:delta_omega_frac_4} can still be used to qualitatively predict the measured redshifts. When $\varepsilon_\mathrm{hs}=\varepsilon_\mathrm{fin}\rightarrow0^-$ (i.e., $\mathrm{N_e}\approx 10^{20+}\ \mathrm{cm^{-3}}$), the second term in the numerator of eq.~\ref{eq:delta_omega_frac_4} diverges, indicating a strong redshift of the metasurface resonance. However, upon further increasing the carrier density such that $\mathrm{N_e}> 10^{20}\ \mathrm{cm^{-3}}$, the resonance is still redshifted, but the magnitude of redshift decreases as the two terms in the numerator of eq.~\ref{eq:delta_omega_frac_4} compete.

The field enhancement in intensity $|\mathbf{E}|^2$ due to the excitation of the ED resonance in the Ge block is shown in Fig.~\ref{Fig2}b for two representative cases. In the first case, marked by a blue star in Figs.~\ref{Fig2}a and b, we are in the perturbative regime with $0<\varepsilon_\mathrm{hs}<\varepsilon_\mathrm{ini}$. Fig.~\ref{Fig2}b shows the intensity profile of the ED resonance blueshifted to $\lambda_{\mathrm{pert}}=4.19\ \mu\mathrm{m}$ when $\ \mathrm{N_e^{pert}}=1.6\times 10^{19}\ \mathrm{cm}^{-3}$. We choose these parameters to illustrate the perturbative case, since $\varepsilon_{\mathrm{hs}}(\lambda_{\mathrm{pert}},\mathrm{N_e^{pert}})\sim10<\varepsilon_\infty\approx16$. The spatial profile of $|\mathbf{E}|^2$ is quite similar to the unperturbed case when $\varepsilon_\mathrm{hs}=\varepsilon_\mathrm{ini}$ (Fig.~\ref{Fig1}b). The second case, marked by a red star in Figs.~\ref{Fig2}a and b, lies beyond the perturbative regime where $\varepsilon_\mathrm{hs}\ll0$ and the hot spot effectively behaves like a metal. We choose the resonance at $\lambda_{\mathrm{PEC}}=5.09\  \mu\mathrm{m}$ when $\mathrm{N_e^{PEC}}=3.2\times10^{20}\ \mathrm{cm}^{-3}$ and refer to this as the PEC case ($\varepsilon_{\mathrm{hs}}(\lambda_{\mathrm{PEC}},\mathrm{N_e^{PEC}})\sim-165$). In this case, we find that the electric field has been expelled from the hot spot and the spatial profile of $|\mathbf{E}|^2$ is strongly modified. However, the spatial profile across the rest of the Ge block indicates that the same ED resonance has been excited. The transmission spectra of the metasurface for three different values of the FC density at the hot spot $\mathrm{N_e}(\mathbf{r},\mathrm{t})$ are shown in Fig.~\ref{Fig2}c. The black curve shows the metasurface spectrum for the unperturbed case before the pump generates any FC ($\varepsilon_\mathrm{hs}=\varepsilon_\mathrm{ini}$), with the ED resonance at $\lambda_{\mathrm{i}}=4.34\ \mu\mathrm{m}$. The blue curve shows the spectrum for $\mathrm{N_e}=\mathrm{N_e^{pert}}$, where we see significant blueshifting of the ED resonance. The resonance blueshifts to $\lambda_{\mathrm{pert}} = 4.19\ \mu\mathrm{m}$, and the $Q$ factor increases to $Q_\mathrm{{pert}}=318$. However, upon further increasing $\mathrm{N_e}$ while maintaining $\varepsilon_\mathrm{hs}>0$, the quality factor of the resonance decreases, and it vanishes completely in the regime where $\varepsilon_\mathrm{hs}=0$. The red curve shows the spectrum for $\mathrm{N_e}=\mathrm{N_e^{PEC}}$ when the hot spot is metallized and the resonance redshifts significantly to $\lambda_{\mathrm{PEC}}=5.09\ \mu\mathrm{m}$ with $Q_\mathrm{{PEC}}=53$.

An electrostatic model that calculates the effective permittivity of the meta-atom for variable free-electron densities within the hot spot is presented in Section S3 of the SI. The model qualitatively predicts the shifts in the metasurface resonance frequency when the hot spot's relative permittivity is varied between $-\infty<\varepsilon_\mathrm{hs}\leq\varepsilon_\infty$ by calculating the effective capacitance and hence the effective permittivity due to the hot spot inclusion ($\varepsilon_\mathrm{hs}(\mathrm{N_e}(\mathbf{r},\mathrm{t}))$) inside a capacitive meta-atom when a constant potential difference is applied across the meta-atom. 

\section{Modeling Time Interfaces in Semiconductors Using Free Carrier Generation} \label{sec:time_domain}
Locally generating FCs within the nanostructures of a metasurface enables dynamic tuning of the effective permittivity and resonance frequency of the metasurface. This makes LFCG in an all-dielectric metasurface a promising platform for time-varying photonics at mid-IR wavelengths. To understand the optical effects of LFCG in the hot spot, we modeled each meta-atom as a time-varying Drude-Lorentz medium and derived the wave equation for the time-varying metasurface. Specifically, we study the impact of resonance shifting via controlled LFCG in the hot spot to theoretically demonstrate that the metasurface resonance can be tuned on the order of an optical cycle of the mid-IR probe (i.e., the metasurface-guided wave, or MGW) by an optical pump pulse, thereby realizing a sharp TI. In subsequent sections, we study in detail the case in which LFCG leads to a redshift of the mid-IR probe MGW.

\subsection{Laser-Induced Free Carrier Generation}
\label{sec:IIIA}
Upon irradiation of an intense laser pulse on a solid, FCs can be generated via the photoionization process. We use the Keldysh model \cite{Gruzdev2014SPIE} to calculate the photoionization rate $\rm{R_{PI}}$ and obtain the FC density ($\mathrm{N_e}$) in Ge as
\begin{equation}
    \frac{\partial \mathrm{N_e}}{\partial \mathrm{t}} = \mathrm{R_{PI}(I[t])}-\frac{\mathrm{N_e}}{\tau_\mathrm{e-h}}
    \label{Keldysh}
\end{equation}
where $\tau_\mathrm{e-h}$ is the electron-hole recombination time, and the partial time derivative indicates that $\rm{N_e}$ has both spatial and time dependence. We use a pump pulse centered at $\lambda_\mathrm{p}=\mathrm{1580\ nm}$, with pulse width $\tau_\mathrm{p}=$ 8 fs, focused at the hot spot for LFCG, which modifies $\varepsilon_{\rm{hs}}$. Near-IR ultrafast pulses have been experimentally realized \cite{MaxPlanck2021}, and are essential for rapid FC generation to produce a sharp TI. The pump intensity incident at the hot spot is $\mathrm{I}[\mathrm{t}]=\mathrm{2I_{avg}cos^2(k_{p}( z+\mathrm{ct}))}$ $\mathrm{e}^{-2(\mathrm{t}-\tau_{1})^2/\tau_{\mathrm{p}}^2}$, where $\mathrm{k_{p}}=2\pi/\lambda_\mathrm{p}$, and we choose $\mathrm{I_{avg}=2.075}$ $\rm{TW/cm^2}$ and $\tau_{1}=3\ \mathrm{ps}$. $\tau_\mathrm{e-h}$ in Ge is on the order of hundreds of picoseconds and thus is much longer than the time period and pulse width of the optical pump and the mid-IR MGWs \cite{Ge_dynamics2017}. Therefore, the FC population in the hot spot is maintained for a sufficiently long time, and we restrict our analysis to the sharp TI between the two phases of the hot spot, i.e., the unionized phase (before the TI) and the ionized phase (after the TI). Further details on the calculation of $\mathrm{N_e}$ using the Keldysh model are provided in Section S4 of the SI. The Keldysh model for photoionization by an optical field accounts for both the tunneling ionization and the multiphoton absorption. For the material comprising the metasurface, Ge (band gap, $\Delta\approx 0.8$ eV, and electron's reduced mass, $\mathrm{m_{e}}=0.041\ \mathrm{m_0}$) \cite{van2011book}, and the given pump pulse parameters, the Keldysh parameter $\gamma_\mathrm{p}= {\mathrm{k_pc}\sqrt{\mathrm{m_e}\Delta}}/{\mathrm{eF}}\approx 0.547$ (average electric field strength $\mathrm{F}=\sqrt{{2 \mathrm{I_{avg}}}/\mathrm{c}\epsilon_0\mathrm{n_p}^2}$, $\mathrm{n_p}=4.2$ is the Ge refractive index at the pump wavelength) \cite{Gruzdev2014SPIE, Amotchkina2020}. Since $\gamma_{\mathrm{p}}\ll 1$, tunneling ionization is the dominant mechanism \cite{Boroumand2022} and the newly formed free electrons possess negligible kinetic energy.

\subsection{Time-Varying Dispersive Medium}
\label{subsec:TVDLM}
We model Ge as a Drude-Lorentz dispersive medium to understand the optical effects of FC generation at the hot spot by an intense pump pulse. The displacement field $\mathbf{D}$ and the polarization current $\mathbf{J}_\mathrm{e}$ in a Drude-Lorentz medium are $    \mathbf{D} = \epsilon_0 \varepsilon_\infty \mathbf{E} + \mathbf{P}_\mathrm{e}$, where, $\mathbf{J}_\mathrm{e}=\partial_\mathrm{t}\mathbf{P}_\mathrm{e}
$ where $\mathbf{P}_\mathrm{e}$ is the density of the FC-induced dipole moment. The polarization current, $\mathbf{J}_\mathrm{e}$, is formed as a result of the interaction of MGW fields with FCs. We consider a medium with pre-existing free electron density $\mathrm{N^{(1)}_{e}}$ and a time-dependent density of electrons created by tunneling ionization $\mathrm{N^{(2)}_{e}(t)}$ such that the total FC density in the medium is $\mathrm{N_e(t)=N^{(1)}_e+N^{(2)}_e(t)}$. Under the two-fluid approximation for two different electronic populations, where we consider only the electronic motion in a neutral plasma, the polarization current density is given by, $ \mathbf{J}_\mathrm{e}(\mathrm{t})= \mathbf{J}^\mathrm{(1)}_\mathrm{e}(\mathrm{t}) + \mathbf{J}^\mathrm{(2)}_\mathrm{e}(\mathrm{t})=-\mathrm{eN^{(1)}_e }\mathbf{v}^{(1)}_\mathrm{e}(\mathrm{t})-\mathrm{e}\int_{-\infty}^\mathrm{t}\partial_\mathrm{t'}\mathrm{N^{(2)}_e}\mathbf{v}^{(2)}_\mathrm{e}(\mathrm{t,t'})\mathrm{dt'}$, where $\mathbf{J}^{(1)}_\mathrm{e}$ and $\mathbf{J}^{(2)}_\mathrm{e}$ are the electron current densities due to the pre-existing and newly created free electrons, respectively. $\mathbf{v}^{(1)}_\mathrm{e}(\mathrm{t})$ is the velocity of a pre-existing electron, and $\mathbf{v}^{(2)}_\mathrm{e}(\mathrm{t,t'})$ is the velocity at time $\mathrm{t}$ of an electron created at time $\mathrm{t'}$ (where $\mathrm{t>t'}$) \cite{Dawson1993}. These velocities, due to acceleration by the electric field of an electromagnetic wave, are calculated as $\mathbf{v}^{(1)}_\mathrm{e}(\mathrm{t})=-({\mathrm{e}}/{\mathrm{m_e}})\int_\mathrm{-\infty}^\mathrm{t}\mathbf{E}(\mathrm{t''}) \mathrm{dt''}=({\mathrm{e}}/{\mathrm{m_e}})\mathbf{A}(\mathrm{t}),\ \mathbf{v}^{(2)}_\mathrm{e}(\mathrm{t,t'})=-({\mathrm{e}}/{\mathrm{m_e}})\int_\mathrm{t'}^\mathrm{t}\mathbf{E}(\mathrm{t''}) \mathrm{dt''}=({\mathrm{e}}/{\mathrm{m_e}})[\mathbf{A}(\mathrm{t})-\mathbf{A}(\mathrm{t'})]$, where $\mathbf{A}$ is the magnetic vector potential and we have used $\mathbf{E}=-\partial_\mathrm{t}\mathbf{A}$. Note that when calculating $\mathbf{J}_\mathrm{e}^{(2)}$, we only consider the acceleration of the FCs after their creation at time $\rm{t'}$. Here we neglect the non-parabolicity of the electronic bands and assume the same effective mass $\mathrm{m_e}$ for all electrons in the conduction band of Ge. The sum of the two polarization current densities due to the pre-existing carriers and the new carriers generated by LFCG due to the pump pulse is given as
\begin{equation}
    \mathbf{J}^{(1)}_\mathrm{e}+\mathbf{J}^{(2)}_\mathrm{e} =\frac{\partial \mathbf{P}_\mathrm{e}}{\partial \mathrm{t}}= -\frac{\mathrm{e^2N^{(1)}_e}}{\mathrm{m_e}}\mathbf{A}\mathrm{(t)}-\frac{\mathrm{e^2 N^{(2)}_e(t)}}{\mathrm{m_e}}\mathbf{A}\mathrm{(t)}+\frac{\mathrm{e^2}}{\mathrm{m_e}}\int_{-\infty}^\mathrm{t}\frac{\partial\mathrm{N^{(2)}_e}}{\partial\mathrm{t'}}\mathbf{A}(\mathrm{t'})\mathrm{dt'}
\label{eq:J_e}
\end{equation}
Calculating the first-order time derivative of $\mathbf{J}_\mathrm{e}$ from eq.~(\ref{eq:J_e}), yields
\begin{equation}
\begin{aligned}
    \frac{\partial\mathbf{J}_\mathrm{e}}{\partial\mathrm{t}}=\frac{\partial^2\mathbf{P}_\mathrm{e}}{\partial\mathrm{t}^2}=\frac{\mathrm{e^2}}{\mathrm{m_e}}\left[\mathrm{N^{(1)}_e}+\mathrm{N^{(2)}_e(t)}\right]\mathbf{E}=\frac{\mathrm{e^2 N_e(t)}}{\mathrm{m_e}}\mathbf{E}=\epsilon_0\omega_\mathrm{p}^2(\mathrm{t})\mathbf{E}
\end{aligned}
\label{eq:Pe_2}
\end{equation}
Here, we emphasize that $\partial_\mathrm{t}\mathbf{J}_\mathrm{e}$ is independent of $\partial_\mathrm{t}\mathrm{N_e}$. This implies that the free electrons created at time 't' do not contribute to the current density at the same instant because the electrons are created at rest. Now, starting from Ampere's law, we formulate the wave equation in terms of $\mathbf{A}$ as
\begin{equation}
   {\nabla} \times ({\nabla} \times \mathbf{A}) + \frac{\varepsilon_\infty}{\mathrm{c^2}} \frac{\partial^2 \mathbf{A}}{\partial \mathrm{t}^2} - \mu_0 \frac{\partial \mathbf{P}_\mathrm{e}}{\partial \mathrm{t}}=0; \ \mathbf{H}=\frac{1}{\mu_0}(\nabla\times \mathbf{A})
   \label{eq:COMSOL Wave Equation}
\end{equation}
COMSOL's time-domain solver uses Eq.~(\ref{eq:COMSOL Wave Equation}) in combination with Eq.~(\ref{eq:Pe_2}) to simulate the interaction between an MGW and FCs at the hot spot. We numerically solve Eq.~(\ref{Keldysh}) to obtain the temporal evolution of $\mathrm{N_e}$ for a pump pulse and use it to model a sharp TI for the MGW. During the TI, FCs created at rest are accelerated by the electromagnetic field, thereby generating DC currents in the system. These constant currents establish a quasistatic magnetic field ($\mathbf{H}_\mathrm{s}$) and consequently a quasistatic magnetic vector potential ($\mathbf{A}_\mathrm{s}$) in the system. The quasistatic magnetic field (QS) mode is a zero-frequency mode that is temporally invariant but spatially varying. After the TI ($\mathrm{t>t_{TI}}$), under the quasi-monochromatic wave approximation, we can write the total magnetic vector potential as $\mathbf{A}(\mathbf{r},\mathrm{t})=\mathbf{A}_\mathrm{s}(\mathbf{r})+\mathbf{A}_\mathrm{t}(\mathbf{r},\mathrm{t})+\mathrm{c.c.}$, where $\mathbf{A}_\mathrm{t}(\mathbf{r},\mathrm{t})=\mathbf{A}_\mathrm{t0}(\mathbf{r})\mathrm{e}^{-\mathrm{i}\omega_\mathrm{f}\mathrm{t}}$ denotes the amplitude of the time-varying or AC component of the magnetic vector potential and $\omega_\mathrm{f}$ is the frequency of the EM wave after TI. In general, for $\mathrm{t>t_{TI}}$, using the ansatz above for $\mathbf{A}(\mathbf{r},\mathrm{t})$ in eqs.~(\ref{eq:J_e}, \ref{eq:COMSOL Wave Equation}), and taking only the time-independent components (see Section S7 in the SI), the amplitude of the QS mode at the hot spot $\mathbf{A}(\mathbf{r})$ satisfies
\begin{equation}
    \mathbf{H}_\mathrm{s}(\mathbf{r})=\frac{1}{\mu_0}\nabla\times\mathbf{A}_\mathrm{s}(\mathbf{r});\ \ (\omega_\mathrm{p}^2-\mathrm{c^2}\nabla^2)\mathbf{A}_\mathrm{s}(\mathbf{r})=\frac{\mathrm{e^2}}{\epsilon_0\mathrm{m_e}}\int_{-\infty}^\mathrm{t}\frac{\partial\mathrm{N^{(2)}_e}}{\partial\mathrm{t'}}\mathbf{A}(\mathbf{r},\mathrm{t'})\mathrm{dt'}
    \label{eq:A_s}
\end{equation}
The net radiation from the QS mode is zero ($\mathbf{E}_\mathrm{s}=-\partial_\mathrm{t}\mathbf{A}_\mathrm{s}=0$). This nonradiative zero-frequency mode consumes a finite fraction of the total energy of the MGW, which is stored in the FCs' DC motion. In the next section, we study the energy dynamics in a time-varying Drude-Lorentz medium. 

\subsection{Energy Relations}
\label{subsec:Energy}
In this section, starting from Poynting's theorem, we derive an analytic expression for the total energy density in a time-varying Drude-Lorentz medium.  Then, we numerically demonstrate that the total energy in a dispersive system is conserved across a TI. During simulations, electromagnetic radiation escapes the simulation domain through the scattering boundaries. From Poynting's theorem, the rate of change of the total energy density can be written as (see Section S6 in the SI), $\partial_\mathrm{t} \mathrm{U}=-\nabla \cdot(\mathbf{E}\times\mathbf{H})=0.5*\partial_\mathrm{t}\left(\epsilon_0\varepsilon_\infty|\mathbf{E}|^2+\mu_0 |\mathbf{H}|^2\right)+\mathbf{E}\cdot\partial_\mathrm{t}\mathbf{P}_\mathrm{e}$. Using $\mathbf{A}(\mathbf{r},\mathrm{t})=\mathbf{A}_\mathrm{s}(\mathbf{r})+\mathbf{A}_\mathrm{t}(\mathbf{r},\mathrm{t})+\mathrm{c.c.}$ in conjunction with the expressions for current given in eq.~(\ref{eq:J_e}), we can separate the time-dependent electromagnetic energy density terms and the quasistatic current driven DC terms for $\mathrm{t>t_{TI}}$. The electromagnetic energy density is $\mathrm{U_{EM}}=0.5*(\epsilon_0\varepsilon_\infty|\mathbf{E}|^2 + \mu_0 |\mathbf{H}|^2 +\epsilon_0\omega_\mathrm{p}^2 (\mathrm{t})|\mathbf{A}_\mathrm{t}|^2)$ and the residual or DC in the system due to the QS mode is $\mathbf{J}_\mathrm{{e,s}}=-\left({\mathrm{e}^2{\mathrm{N_e(t)}}}/{\mathrm{m_e}}\right)\mathbf{A}_\mathrm{s}(\mathbf{r})+({\mathrm{e}^2}/{\mathrm{m_e}})\int_{-\infty}^\mathrm{t}\partial_\mathrm{t'}\mathrm{N^{(2)}_e}\mathbf{A}(\mathbf{r},\mathrm{t'})\mathrm{dt'}$. Then, using these terms, we obtain $-\partial_\mathrm{t}\mathrm{U_{EM}}=\nabla \cdot(\mathbf{E}\times\mathbf{H})+\mathbf{E}\cdot\mathbf{J}_\mathrm{e,s}$. We observe that the rate of change of $\rm{U_{EM}}$ after the TI is equal to the sum of the electromagnetic flux leaving the system and the rate of work done on the static currents created in the system during a TI. We emphasize that $\mathbf{J}_\mathrm{e,s}$ is a constant current density related to the DC motion of the FCs that is created during a TI and persists even after the departure of the electromagnetic waves. Finally, integrating the equation for total energy (see Section S6 of the SI for detailed derivations), yields
\begin{equation}
\begin{aligned}
    \mathrm{U}=\frac{1}{2}\epsilon_0|\mathbf{E}|^2+\frac{1}{2}\mu_0 |\mathbf{H}|^2+\frac{1}{2}\epsilon_0&(\varepsilon_\infty-1)|\mathbf{E}|^2+\frac{1}{2}\epsilon_0\omega^2_\mathrm{p}|\mathbf{A}|^2\\ +\int_{-\infty}^{\mathrm{t}}\frac{\partial\mathrm{N^{(2)}_e}}{\partial\mathrm{t'}}\frac{|\mathrm{e}\mathbf{A}(\mathrm{t'})|^2}{2\mathrm{m_e}}\mathrm{dt'}-&\mathbf{A}(\mathrm{t})\cdot\left(\frac{\mathrm{e^2}}{\mathrm{m_e}}\int_{-\infty}^\mathrm{t}\frac{\partial\mathrm{N^{(2)}_e}}{\partial\mathrm{t'}}\mathbf{A}(\mathrm{t'})\mathrm{dt'}\right)
\end{aligned}
\label{Energy}
\end{equation}
The expression above for the total energy density holds at all times, including the time before, during, and after the time interface. The first (second) term on the RHS of eq.(\ref{Energy}) represents the energy density stored in the form of electric (magnetic) fields. In contrast, the third term represents the potential energy density stored in the bound electrons. The sum of the first three terms is labeled $\mathrm{U_{Field}}$, while the sum of the remaining terms represents the energy density stored in the FCs and is labeled $\mathrm{U_{Carrier}}$. In the next section, we use eq.~(\ref{Energy}) to calculate and compare the total energies for the cases with a TI and without a TI. 
\begin{figure}[t]
\centering
    \includegraphics[width=155mm]{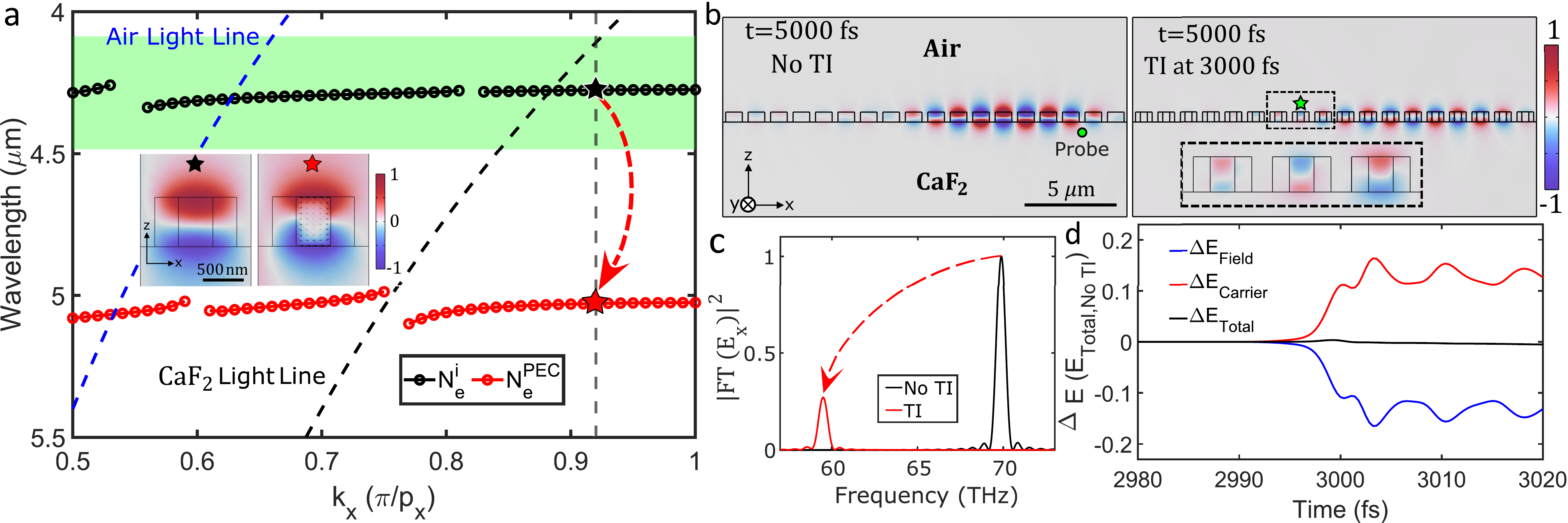}
    \caption{$\bf{a.}$ Dispersion plot of the ED mode; green shaded region indicates the excitation source (dipole) spectrum. Inset: $\mathrm{H_y}$ field profile of the ED mode (in the $x$-$z$ midplane) for the black/red star markers. $\bf{b.}$ Normalized $\mathrm{H_y}$ profile in the $x$-$z$ mid-plane of the simulation domain on the left (right) shows the MGW propagating with group velocity $\mathrm{v^i_g \approx 0.007c}$ ($\mathrm{v^{PEC}_g \approx 0.009c}$) without a TI (with a TI) at t=5000 fs; the rectangular region inside each meta-atom represents the cylindrical hot spot of radius 210 nm. The light green star indicates the meta-atom, where the magnetic field is shown in Fig.~\ref{Fig4}. $\bf{c.}$ $\mathrm{|FT(E_x)|^2}$ recorded at the probe after the TI for the two cases: $\mathrm{N^{(2)}_e}=\mathrm{N^{i}_e=0\ cm^{-3}}$ (No TI, black line), and $\mathrm{N^{(2)}_e=N^{PEC}_e=3.2\times10^{20}\ cm^{-3}}$ (TI, red line). $\bf{d.}$ Time evolution of $\Delta\mathrm{E_{Field}}$: blue line, $\Delta \mathrm{E_{Carrier}}$: red line, and$\Delta \mathrm{E_{Total}}$: black line.}
    \label{Fig3}
\end{figure}

\section{Impact of a Time Interface in a metasurface}
\label{sec:TI_in_metasurface}
\subsection{Temporal Scattering of MGWs}
\label{subsec:IV_A}
The TI is created by FC generation in the hot spot under intense pump-pulse illumination, resulting in a time-varying metasurface. A MGW propagating along the metasurface is temporally scattered by the TI created by this time-varying metasurface. Metasurface eigenmode and TI simulations are performed using COMSOL Multiphysics. The metasurface dispersion plot for the ED mode is shown in Fig.~\ref{Fig3}a for the unperturbed (black line) and perturbed (red line) cases. Although the ED mode of the unperturbed metasurface leaks into the $\mathrm{CaF_2}$ substrate for most of the Brillouin zone, it drops below the light line at high $\rm{k_x}$. It becomes an ED MGW (decaying in the z direction away from the metasurface) near the edge of the Brillouin zone ($\mathrm{|k_x|}\gtrapprox 0.89\cdot\pi/\mathrm{p_x}$). We excite an ED MGW, propagating in the $\mathrm{x}$ direction with $\mathrm{k_x}=0.92\pi/\mathrm{p_x}$, $\mathrm{\ k_y=k_z=0}$ (dashed gray line in Fig.~\ref{Fig3}a) $\varepsilon_\mathrm{hs}=\varepsilon_\mathrm{ini}$. The MGW has a larger wave vector than the free-space light, $\mathrm{|k_x|}>2\pi/\lambda_0$, where $\lambda_0=4.27\ \mu \mathrm{m}$ is the central wavelength of the mid-IR probe, which excites the ED MGW. Thus, we used an array of phased electric dipoles to excite the MGW. These dipoles, as well as the simulation domain, are shown in Fig.~S5a in the SI. The dipole moment of the $n$-th dipole (at $x_n=nd_x,\ n$=1,2,3,...) at time t, $\textbf{\textit{p}}_{\mathrm{dipole}}(x_n,\mathrm{t})=\textbf{\textit{p}}_0\mathrm{cos(k_x}x_n-\omega_{0}\mathrm{t})\mathrm{e}^{-(\mathrm{t}-\tau_{2})^2/\tau_{0}^2} \mathrm{\hat z}$ (where $\omega_{0}={2\pi \mathrm{c}}/{\lambda_{0}}$, $\textbf{\textit{p}}_0=1$ C-m $\tau_{2}=300$ fs and $\tau_{0}=200$ fs). The green-shaded region in Fig.~\ref{Fig3}a shows the spectrum of the dipolar excitation. The phased dipole array is essential to efficiently excite the MGW at a specific $\mathrm{k_x}$, as the dispersion line for the ED MGW is extremely flat. 

Alterations in the constitutive materials of the metasurface over time (e.g., via controlled LFCG at the hot spot by a pump pulse) modify the metasurface, resulting in a time-varying medium for the MGW. For the pump pulse parameters given in Section \ref{sec:IIIA}, we calculated the FC concentration in the hot spot using eq.~(\ref{Keldysh}) and the transition time in which $\mathrm{N_e}$ changes from $\mathrm{N_e^i}$ to $\mathrm{N_e^{PEC}}$ is $\mathrm{T_{trans} \approx}\ 16\ \mathrm{fs} \gtrsim \mathrm{T}_{0}=2\pi/\omega_0 \approx 14.27$ fs. The transition from $\mathrm{N_e^{i}}$ to $\mathrm{N_e^{PEC}}$ occurs in slightly over an optical cycle of the MGW, thus creating a moderately sharp TI for the propagating MGW. We compare the $\mathrm{H_y}$ field profiles at 5000 fs in the cases with and without a TI. For the case without a TI, the MGW crosses approximately 8 meta-atom unit cells (see Fig. \ref{Fig3}b) from left to right in the simulation domain, propagating at a group velocity of $\mathrm{v^i_g \approx 0.007c}$, where c is the speed of light in vacuum. We compare this case with that of a TI at 3000 fs. In 3000 fs, the unperturbed MGW propagates across 5 meta-atoms. However, at 5000 fs in the case with a TI, we observe that the MGW crosses one extra meta-atom propagating at an increased group velocity of $\mathrm{v^{PEC}_g \approx 0.009c}$ in comparison to the unperturbed MGW, thus crossing a total of 9 meta-atoms. To clearly view the magnetic field rectified during the TI, we show an enlarged view of some of the meta-atoms enclosed in a dashed black box in the right-field plot of Fig. \ref{Fig3}b. The enclosed meta-atoms clearly exhibit a rectified magnetic field in the hot spot region. Furthermore, we provide evidence of temporal scattering by showing that the MGW frequency redshifts across the TI \cite{Filiberto2024}. The x component of the electric field ($\mathrm{E_x}$) is recorded at the probe location (shown as a yellow circle in the left panel in Fig.~\ref{Fig3}b). Fourier transforms of $\mathrm{E_x}$ for the two cases, with a TI at 3000 fs ($\mathrm{|FT(\mathrm{E}_\mathrm{PEC})|^2}$) and without a TI ($\mathrm{|FT(\mathrm{E}_\mathrm{i})|^2}$), are shown in Fig.~\ref{Fig3}c. In both cases, we focus exclusively on the electric field measured after 3000 fs. For the case without a TI, we observed a peak in the spectrum (black line in Fig.~\ref{Fig3}c) at 70.1 THz ($\approx$ 4.27 $\mu$m) corresponding to the ED resonance of the metasurface. Now, in the presence of a sharp TI (red line in Fig.~\ref{Fig3}c), the spectrum of the recorded field exhibits a clearly red-shifted ED peak at $\sim$59.4 THz ($\approx$ 5.05 $\mu$m). Thus, we see that the TI due to $\mathrm{N_e}$ changing from $\mathrm{N_e^i}\to\mathrm{N_e^{PEC}}$ results in the redshifting of the MGW by exactly the amount predicted by the eigenmode dispersions for $\mathrm{N_e^i}$ and $\mathrm{N_e^{PEC}}$ FC densities (dashed red arrow in Fig.~\ref{Fig3}a).

\subsection{Energy Dynamics Across a Time Interface}
\label{subsec:IV_b}
We consider pre-existing FC density $\mathrm{N^{(1)}_e}=0$ for the TI simulations. We integrate the expressions for the energy densities from eq.~(\ref{Energy}) over the simulation domain to calculate the time evolution of the energy contributions of the different terms (that is, $\mathrm{E_{Field}=\int U_{Field}dV}$, and $\mathrm{E_{Carrier}=\int U_{Carrier}dV}$). We calculate the difference in $\mathrm{E_{Field}}$, $\mathrm{E_{Carrier}}$, and also the total energy of the system ($\mathrm{E_{Total}=\int UdV}$) for the cases with and without a TI ($\Delta \mathrm{E_{Total}}=\Delta \mathrm{E_{Field}}+\Delta \mathrm{E_{Carrier}};$ $\Delta \mathrm{E=E(TI)-E(No\ TI)}$ and plot these in Fig.~\ref{Fig3}d. We observe that when the TI is introduced, $\Delta \mathrm{E_{Field}}$ decreases, while $\Delta \mathrm{E_{Carrier}}$ increases by an equal magnitude. Note that $\Delta \mathrm{E_{Total}}$ drops slightly below the zero line after the TI. This can be attributed to the simulation domain not being a closed system, as it contains scattering boundaries. Since the group velocity increases and the quality factor of the MGW decreases after TI, the energy in the system escapes the system more rapidly through the boundaries. In any case, ignoring these small contributions to the total energy, our simulations indicate that the system's total energy in the presence of TI (created via LFCG) is the same as in its absence, i.e, U(no TI)$|_\mathrm{t>t_{TI}}$ = U(TI)$|_\mathrm{t>t_{TI}}$ where $\rm{t_{TI}=3000\ fs}$. Now, for the case of no TI, we use the unperturbed fields ($\mathbf{E}_0$, $\mathbf{H}_0$, $\omega_0$) to calculate the energy density U(no TI)$|_\mathrm{t>t_{TI}}$ after the TI, which, time averaged over an optical cycle, can be written as
\begin{equation}
    \mathrm{U(no\ TI)}|_\mathrm{t>t_{TI}}=\mathrm{U_{EM}}\ (\mathrm{no\ TI})=\frac{1}{2}\epsilon_0\varepsilon_\infty\langle|\mathbf{E}_0|^2\rangle + \frac{1}{2}\mu_0 \langle|\mathbf{H}_0|^2\rangle+\frac{1}{2}\epsilon_0\left(\frac{\omega_\mathrm{p0}^2}{\omega_0^2}\right)\langle|\mathbf{E}_0|^2\rangle
    \label{eq:U_no_TI}
\end{equation}
where $\omega_\mathrm{p0}$ and $\mathrm{U_{EM}}\ (\mathrm{no\ TI})$ are the plasma frequency and the electromagnetic energy density for the case without TI, respectively. Here, \enquote{$\langle\ \rangle$} denotes the time-averaged quantity over an optical cycle of the wave's frequency. However, for the case with the TI, to calculate the time-averaged total energy density, we use $\mathbf{A}(\mathbf{r},\mathrm{t})=\mathbf{A}_\mathrm{s}(\mathbf{r})+\mathbf{A}_\mathrm{t}(\mathbf{r},\mathrm{t})+\mathrm{c.c.}$ in eq.~(\ref{Energy}) to separate the energy density in propagating electromagnetic fields from the energy density in the QS mode such that $\mathrm{U(TI)}|_\mathrm{t>t_{TI}}=\mathrm{U_{EM}}\ (\mathrm{TI})+\mathrm{U_{QS}}$. 
\begin{equation}
\begin{gathered}
    \mathrm{U_{EM}}\ (\mathrm{TI})=\frac{1}{2}\epsilon_0\varepsilon_\infty\langle|\mathbf{E}_\mathrm{t}|^2\rangle + \frac{1}{2}\mu_0 \langle|\mathbf{H}_\mathrm{t}|^2\rangle+\frac{1}{2}\epsilon_0\left(\frac{\omega_\mathrm{p}^2}{\omega_\mathrm{f}^2}\right)\langle|\mathbf{E}_\mathrm{t}|^2\rangle
    \label{eq:U_EM_TI}
    \end{gathered}
\end{equation}
 Here, $\mathrm{U_{EM}}\ (\mathrm{TI})$ is the electromagnetic energy density for the case of TI. The energy density in the hot spot due to the QS mode of the system can be expressed using the quasistatic magnetic field ($\mathbf{H}_\mathrm{s}$) and the quasistatic magnetic vector potential ($\mathbf{A}_\mathrm{s}$) derived in eq.~(\ref{eq:A_s}) as
\begin{equation}
    \mathrm{U_{QS}}=\frac{1}{2}\mu_0 |\mathbf{H}_\mathrm{s}|^2+\frac{1}{2}\epsilon_0\omega^2_\mathrm{p}|\mathbf{A}_\mathrm{s}|^2 +\int_{-\infty}^{\mathrm{t}}\frac{\partial\mathrm{N^{(2)}_e}}{\partial\mathrm{t'}}\frac{|\mathrm{e}\mathbf{A}(\mathrm{t'})|^2}{2\mathrm{m_e}}\mathrm{dt'}-\mathbf{A}_\mathrm{s}\cdot\left(\frac{\mathrm{e^2}}{\mathrm{m_e}}\int_{-\infty}^\mathrm{t}\frac{\partial\mathrm{N^{(2)}_e}}{\partial\mathrm{t'}}\mathbf{A}(\mathrm{t'})\mathrm{dt'}\right)
    \label{eq:U_QS}
\end{equation}
From eqs.~(\ref{eq:U_no_TI},\ref{eq:U_EM_TI},\ref{eq:U_QS}), it follows that in the case of TI, the electromagnetic energy is depleted by the same amount as the energy expended to create DC currents, $\mathrm{U_{EM}\ (TI)}-\mathrm{U_{EM}\ (no\ TI)}=\Delta\mathrm{U_{EM}}=-\mathrm{U_{QS}}$. Numerically, we find $\mathrm{U_{QS}>0}$, which implies that the electromagnetic energy density in the MGW decreases, while the energy density in the QS mode increases across the TI.

\subsection{Nanoscale Magnetization Using Rapid FC Generation}\label{subsec:IV_c}

The MGW electric field accelerates the electrons generated at rest within the hot spot during the TI. Once they acquire a non-zero velocity, they are subject to the magnetic field of the mid-IR MGW present in the hot spot. After the TI, the large number of FCs in the hot spot screen the MGW electric field. Consequently, the displacement currents in the hot spot volume before the TI due to the AC electric field of the MGW are replaced by real electric currents because of the motion of the FCs in the hot spot after the TI. These circulating currents within the hot spot produce a quasistatic magnetic field obtained by rectifying the AC magnetic field of the MGW during the TI. The acceleration of the FCs that are created during the TI by the mid-IR MGW is driven by the magnetic vector potential, $\mathbf{A}\mathrm{(t)}$, of the MGW. The long wavelength mid-IR MGW is better able to accelerate the electrons because $\mathbf{A}\mathrm{(t)}$ scales linearly with the wavelength. Additionally, the energy of the longer-wavelength photon in the mid-IR MGW is below the Ge bandgap, thereby eliminating the possibility of single-photon absorption and, consequently, FC generation. Another process that can generate FC via the mid-IR MGW is multiphoton absorption. However, the multiphoton absorption probability of a mid-IR MGW is drastically reduced because the Keldysh parameter for the MGW, which is approximately the ratio of the TI duration to the MGW optical cycle, is less than one \cite{Maxim2021NComms}. The rectification of the magnetic field of an electromagnetic wave upon the rapid creation of a plasma around a portion of the wave was studied several decades ago \cite{Mori_1988_B_Field}. More recently, the rectification of electric fields in capacitive metasurfaces due to a time interface has been reported \cite{Tretyakov2023Nanophotonics}. The opposite effect has also been predicted, where a static electric field is partially converted to a dynamic radiation field at a time interface \cite{Mencagli2022}. Here, we demonstrate the rectification of the resonantly enhanced magnetic field of an MGW, which is dependent on the full-vector, three-dimensional optical fields of the MGW in the hot spot.

\subsubsection{Quasistatic Magnetic Field}
\label{subsec:IV_c1}
\begin{figure}[t]
\centering
    \includegraphics[width=155mm]{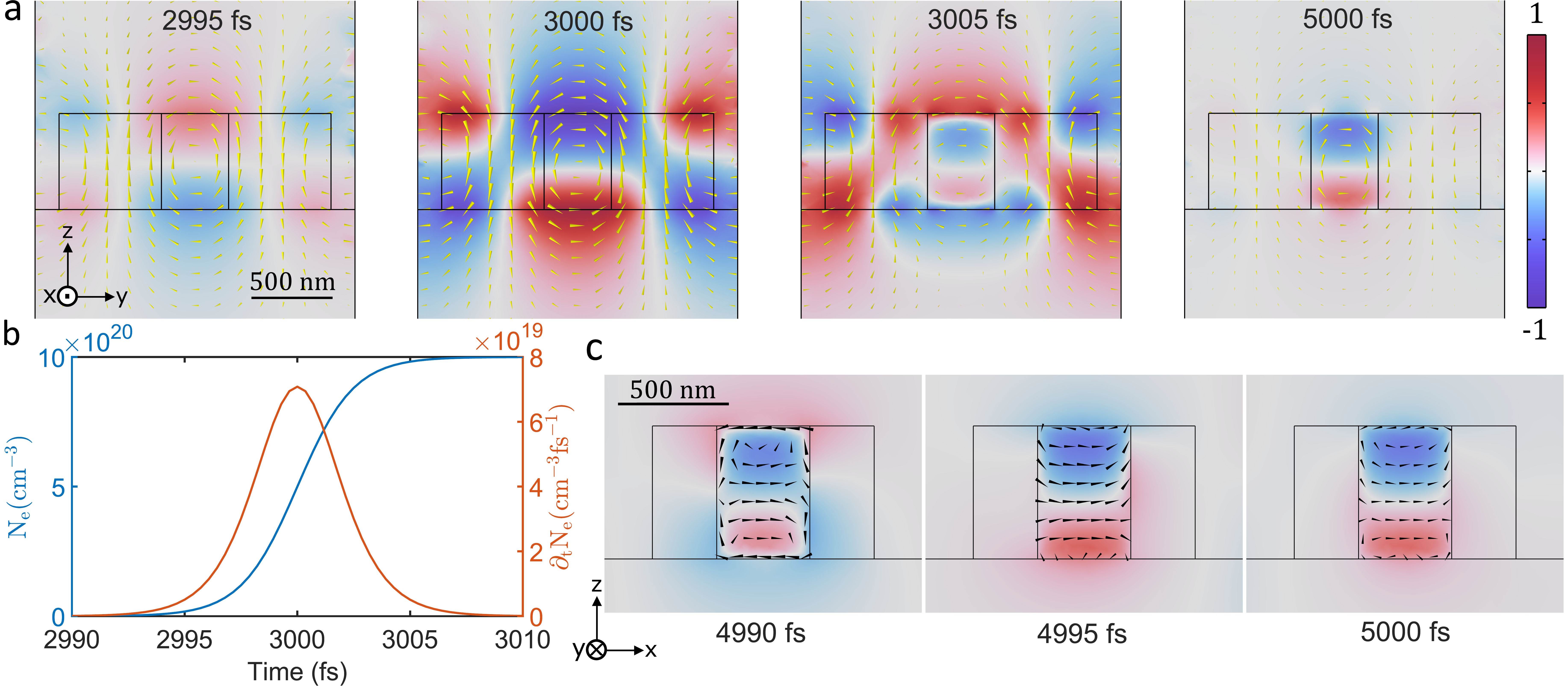}
    \caption{$\bf{a.}$ Normalized $\mathrm{H_y}$ profile in the $y$-$z$ mid-plane of the meta-atom marked using a light green star in Fig.~\ref{Fig3}b, at 2995 fs, 3000 fs, and 3005 fs; the yellow cones represent the magnetic field lines $\bf{b.}$ Time evolution of FC density and its temporal derivative. $\bf{c.}$ Time evolution of the QS mode at three consecutive time steps separated by $\Delta \mathrm{t}=5\ \mathrm{fs}$ in the $x$-$z$ midplane of the same meta-atom; the black cones indicate the FC current density in the hot spot. In all plots, the rectangular region within the meta-atom represents the cylindrical hot spot with radius 210 nm.}
    \label{Fig4}
\end{figure}
The rectification of the magnetic field of the MGW at the TI is shown in Fig. \ref{Fig4}a, which shows snapshots of $\mathrm{H_y}$ in the $y$-$z$ mid-plane at different times inside the meta-atom and its vicinity in air and the substrate. The yellow cones show the direction of the magnetic field. $\mathrm{H_y}$ is normalized to the peak field of steady-state MGW before TI. The first panel of Fig. \ref{Fig4}a shows the magnetic field at 2995 fs, when some FCs are present in the hot spot. The second (third) panel shows the magnetic field of the MGW at 3000 (3005) fs, during the middle (towards the end) of the TI. Compared with the first and second panels, the third panel exhibits marked differences in the spatial structure of the magnetic field. There is a clear discontinuity in the magnetic field close to the boundary between the hot spot and the rest of the unit cell at 3005 fs (smaller rectangle in the center). The magnetic field in the hot spot at 3005 fs retains the spatial pattern of the field at 3000 fs. This is the rectified quasistatic magnetic field produced by the TI (referred to as the QS mode). The fourth panel in Fig.~\ref{Fig4}a shows the magnetic field after approximately 118.5 optical cycles past TI at 5000 fs. We find that, in the absence of losses, the QS mode persists even as the fields from temporally scattered MGW outside the hot spot weaken.

Such magnetic rectification can be understood from eq.~(\ref{eq:A_s}), which shows that the source term for the quasistatic magnetic vector potential $\mathbf{A}_\mathrm{s}$ is a weighted integral of the magnetic vector potential of the MGW during TI where the weights are $\partial_\mathrm{t}\mathrm{N_e^{(2)}}$ (see Fig.~\ref{Fig4}b). The rectified magnetic field in the hot spot at 3005 fs and at 5000 fs maintains a spatial structure resembling the one at 3000 fs, where $\partial_\mathrm{t}\mathrm{N_e^{(2)}}$ is the highest. The rectification of the magnetic field is not perfectly efficient ($\sim 53\%$ efficiency) due to the finite duration of the time interface ($\sim 16$ fs in the case shown in Fig. \ref{Fig4}). An instantaneous TI at a time when the MGW optical fields are at their peak could enable perfectly efficient rectification, but this is an unphysical scenario \cite{Mori_1988_B_Field}. On the other hand, a prolonged TI spanning many optical cycles of the MGW would lead to averaging over multiple cycles, resulting in negligible rectification.

Further evidence of the static nature of the QS mode can be seen in Fig.~\ref{Fig4}c from the $x$-$z$ midplane snapshots that show the field $\mathrm{H_y}$ in three time steps separated by $\Delta \mathrm{t}=5\ \mathrm{fs}=0.3\ \mathrm{T_{f}}$ (where $\mathrm{T_{f}}=16.84\ \mathrm{fs}$ is the optical cycle of the frequency-shifted MGW). Even as the fields outside the hot spot change with time because of the propagation of the temporally scattered MGW at different points in the optical cycle, the QS mode remains unchanged. The electric currents in the hot spot (black cones within the smaller rectangle at the center) replace the displacement currents during the TI and maintain the quasistatic magnetic field, remaining mostly unchanged over the optical cycle of the frequency-shifted MGW. However, there are small alterations near the boundary of the hot spot where the electrons screen the electric field of the temporally scattered MGW outside the hot spot. The absolute magnitude of the rectified magnetic field in the QS mode, $\mathbf{H}_\mathrm{s}$, depends on the optical fields of the MGW. The maximum optical field of MGW that may propagate along the metasurface is, in turn, limited by the laser-induced damage threshold of Ge, which has been measured as $1.7\ \mathrm{TW/cm^2}$ at $3.6\ \mu\mathrm{ m}$ \cite{LIDT_Ge_mid_IR}. We obtain a peak magnetic field amplitude of $\sim2.4\ \mathrm{T}$ when we excite the MGW such that the resonantly enhanced MGW intensity does not exceed $0.17\ \mathrm{TW/cm^2}$ anywhere in the Ge meta-atom. With $\sim53\%$ rectification of the MGW magnetic field for the TI being considered, the peak rectified magnetic field amplitude is $\sim1.272\ \mathrm{T}$. Thus, we find that rectification of a resonantly enhanced AC magnetic field by a TI via LFCG offers a new paradigm for realizing giant, nanoscale magnetization without an external magnetic field.

\subsubsection{Dissipative Losses and Spatially Inhomogeneous FC Generation}
\label{subsec:IV_c2}

In the absence of losses, the rectified magnetic field persists, and its magnitude remains constant. However, the persistence of the QS mode is hindered by the disruption of quasistatic currents arising from electron, ion, and phonon scattering. We consider an arbitrary electronic-scattering frequency, $\gamma^{-1}=100$ fs, which encapsulates contributions from various scattering mechanisms. This leads to a modification of eq.~(\ref{eq:Pe_2}) where a damping term is introduced as ${\partial_\mathrm{t}^2 \mathbf{P}_\mathrm{e}}+\gamma\partial_\mathrm{t}\mathbf{P}_\mathrm{e} = \epsilon_0\omega_\mathrm{p}^2\mathrm{(t)}\mathbf{E}$. Due to these losses, the electronic currents that support the quasistatic magnetic field eventually decay, and the zero-magnetic-field region near the center of the hot spot diffuses until it occupies the entire hot spot. This magnetic diffusion may be modeled using the magnetic diffusion equation, $\partial_\mathrm{t}\mathbf{H}=({1}/{\mu_0\sigma_0})\nabla^2\mathbf{H}=D\ \nabla^2\mathbf{H}$ \cite{jackson2021classical}, where the electronic diffusion constant $D={1}/{\mu_0\sigma_0}$ and $\sigma_0$ is the electrical DC conductivity, which is inversely proportional to $\gamma$. 

\begin{figure*}[t]
\centering
    \includegraphics[width=150mm]{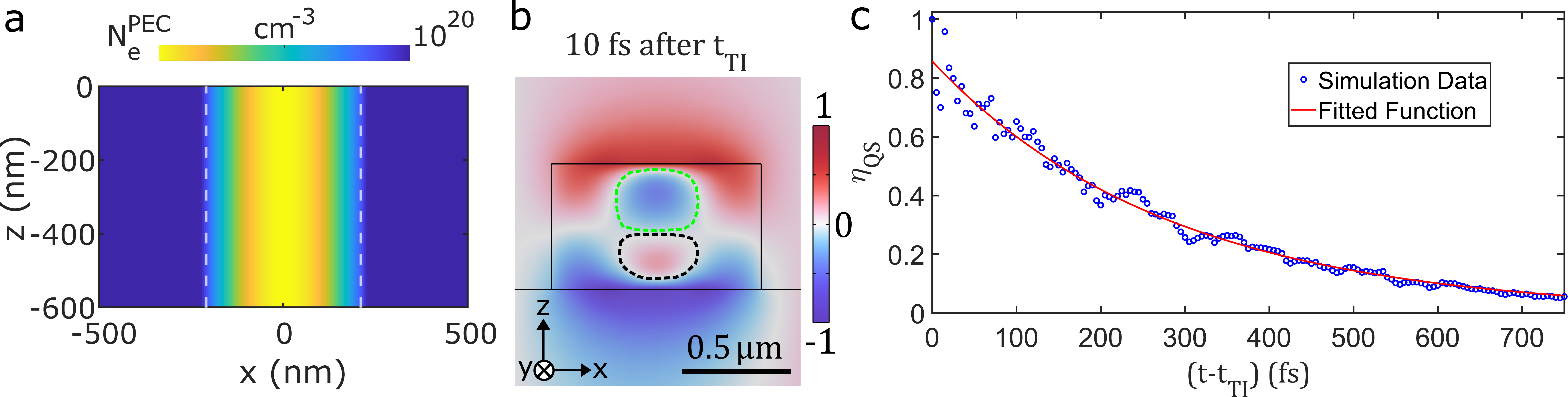}
    \caption{$\bf{a.}$ Spatially inhomogeneous distribution of free carriers in the hot spot obtained from the Keldysh model. $\bf{b.}$ The rectified magnetic field 10 fs after the TI. The dashed polygons enclose the regions inside the hot spot used to calculate $\Delta \mathrm{H_{QS}(t)}$. $\bf{c.}$ $\eta_\mathrm{QS}$ after the time interface, the open circles show the simulation data, while the red line shows the exponential decay function, $\eta_\mathrm{fit}$, fitted to the simulation data.}
    \label{Fig5}
\end{figure*}

Thus far, we had assumed that the hot spot was a homogeneous cylinder with radius $\rm{r_{hs}}$ with constant $\rm{N_e}$. Now, we account for spatially inhomogeneous FC generation arising from the pump pulse's transverse Gaussian intensity profile by using the Keldysh model (see Section S5 of the SI). The spatially inhomogeneous FC density profile obtained is shown in Fig.~\ref{Fig5}a. Furthermore, to quantify the persistence of the QS mode after the TI, we focus on the rectified quasistatic magnetic field and consider the difference between its peak positive and peak negative magnetic fields. We consider two regions in the $x$-$z$ midplane inside the hot spot where the QS mode is dominant over the MGW as shown in Fig.~\ref{Fig5}b and calculate $\Delta\mathrm{H_{QS}}$ by subtracting the minimum magnetic field amplitude in the region enclosed by the dotted green line from the maximum magnetic field amplitude in the region enclosed by the dotted black line. Then, we define a normalized quantity, $\eta_\mathrm{QS}(\mathrm{t})={\Delta\mathrm{H_{QS}(t)}}/{\mathrm{max.}(\Delta\mathrm{H_{QS}(t)})}$.

Thus, lower values of $\eta_\mathrm{QS}$ correspond to smaller amplitudes of the QS mode at the hot spot. We fit the simulation data for $\eta_\mathrm{QS}$ using an exponential decay function that accounts for the diffusion of $\mathbf{H}_\mathrm{s}$ in the presence of losses, $\eta_\mathrm{fit}=\mathrm{C_0}\mathrm{e}^{-\gamma_\mathrm{diff}\mathrm{t}}$. In Fig.~\ref{Fig5}c, which shows the temporal evolution of $\eta_\mathrm{QS}$, the open circles represent the simulation data points and the red line represents the fitted function. From the fitted function, we obtain $\mathrm{C_0}=0.84$, and $\gamma_\mathrm{diff}=3.306\times10^{12}\ \mathrm{rad/s}$ and calculate the diffusion time of the static currents inside the hot spot, $\mathrm{T_{diff}}=\gamma_\mathrm{diff}^{-1}=302.48$ fs. Thus, the rectified field persists for more than three electronic scattering times, extending for nearly 20 optical cycles of the incident MGW. However, the persistence of the QS mode is fundamentally limited by the electron-scattering time, which can be increased by working at lower temperatures as the electron mobility in Ge scales with temperature $T$ as $\mu \propto \gamma^{-1} \propto T^{-1.66}$ \cite{Morin1954Phys.Rev.}. Thus, the 100 fs scattering time we have assumed at room temperature would rise to $\sim$956 fs at 77 K, and consequently, the diffusion time of the QS mode can be extended to $\sim$3 ps or longer by operating at cryogenic temperatures, improving its persistence.

\section{Conclusion}
\label{sec:conclusions}
In conclusion, we have presented a method for tunable shifting of a high-$Q$ resonance in an all-dielectric metasurface resonant in the mid-IR by controlled LFCG in a cylindrical hot-spot region within the meta-atoms. We derived an analytic expression for the frequency shift, taking into account the full vectorial form of the unperturbed electromagnetic fields integrated over the perturbed volume along with the perturbation to the permittivity, that accurately predicts the blueshift of the metasurface resonance due to FC generation in the hot spot in the perturbative regime. We found that, for low FC densities in the hot spot, the resonance blueshifts. However, a further increase in FC density that effectively metalizes the hot spot ($\varepsilon_\mathrm{hs}\ll 0$) results in a strongly red-shifted resonance frequency. Thus, in an experimental setting, carefully tuning the pump intensity to control the local FC density at the hot spot would allow tunable blue- or redshifting of the metasurface resonance.

We have further demonstrated that the shift in the metasurface resonance induced by LFCG provides a platform for studying time-varying photonics in the mid-IR. We studied the effect of LFCG in an array of hot spots at the center of each meta-atom, where we calculated the FC generation due to a pump pulse using the Keldysh model. Modeling each hot spot as a time-dependent Drude-Lorentz medium with rapidly varying plasma density, we developed a theoretical framework to incorporate the effects of rapid free-carrier generation over short but finite time scales into time-domain electromagnetic codes. Implementing a moderately sharp TI using LFCG for a mid-IR MGW, we observed a redshift of the temporally scattered MGWs. Furthermore, using Poynting's Theorem, we identified various contributions to the total energy and derived an analytic expression for the total energy density in a time-varying Drude-Lorentz medium. We demonstrated the re-partitioning of the incident MGW energy into temporally scattered MGW and persistent electric currents and magnetic fields in the domain where LFCG occurs. By comparing cases with and without a TI, we validated that the total energy of the system remains unchanged across the TI and confirmed that LFCG does not act as an energy source or sink for the system. Additionally, after TI, a large, highly localized, and persistent quasistatic magnetic field (QS mode) is observed in the hot spot, which does not radiate but consumes a substantial portion of the electromagnetic energy. Using the TI created by LFCG, we provided evidence of efficient rectification ($\sim53\%$) of the resonantly enhanced AC magnetic field due to the MGW at the hot spot. This shows that a time interface realized via rapid FC generation in an all-dielectric metasurface can serve as a novel platform for efficient conversion of an AC optical field at frequency $\omega$ into a DC optical field. These results significantly enhance and clarify our understanding of energy dynamics in a time-varying medium and of magnetic-field rectification at a TI. We are optimistic that our findings will open new research directions in optically tunable metasurfaces as a platform for time-varying photonics.

\subsection* {Acknowledgments}
The work at Cornell was supported by the University of Dayton Research Institute (UDRI) under contract FA8651-24-F-B013 and by the Office of Naval Research (ONR) under grant no. N00014-21-1-2056, and the Army Research Office (ARO) under the award W911NF2110180. COMSOL time-domain calculations were performed using the computer resources of the Cornell University Laboratory of Plasma Studies, with assistance from Dr. Steven Lantz in the Cornell Center for Advanced Computing.

\bibliography{main} 
\bibliographystyle{ieeetr}

\end{document}


\maketitle

\renewcommand{\thesection}{S\arabic{section}}
\renewcommand{\thefigure}{S\arabic{figure}}
\renewcommand{\thetable}{S\arabic{table}}
\renewcommand{\refname}{SOM References}

\section{Predicting Resonance Frequency Shifts}

The real unperturbed ($\mathbf{{F}}_0$) and perturbed ($\mathbf{F'}$) fields inside the resonator are
\begin{equation}
\mathrm{Re}(\mathbf{{F}}_0)=\mathbf{F}_0(\mathbf{r})\mathrm{e}^{-\mathrm{i\omega_0 t}} \mathrm{+ c.c.};\ \ \ \mathrm{Re}(\mathbf{F}')=\mathbf{F}'(\mathbf{r})\mathrm{e}^{-\mathrm{i\omega t}} \mathrm{+ c.c.}  
\end{equation}

where $\mathbf{F}$ represents the complex fields $ \mathbf{E}, \mathbf{D}, \mathbf{H}$, and $\mathbf{B}$. Here, $\mathbf{F}_0(\mathbf{r})$ and $\mathbf{F}'(\mathbf{r})$ are complex field amplitudes, which are functions of position. Subtracting the scalar product of Maxwell's curl equation for $\mathbf{H}'$ with $\mathbf{E}_0^*$ from the scalar product of the curl equation for $\mathbf{E}'$ with $\mathbf{H}_0^*$ yields the following results.
\begin{equation}
    \mathbf{H}_{0}^* \cdot ({\nabla} \times \mathbf{E}') - \mathbf{E}_0^* \cdot ({\nabla} \times \mathbf{H}') = i\omega \mathbf{B}' \cdot \mathbf{H}_0^* + i\omega \mathbf{D}' \cdot \mathbf{E}_0^*,
    \label{eq:curl_add}
\end{equation}
where $\omega$ is the perturbed complex resonance frequency, and \enquote{$*$} represents the complex conjugate. The LHS of eq. (\ref{eq:curl_add}) can be rewritten using the divergence formula as
\begin{equation}
     \mathbf{H}_{0}^* \cdot ({\nabla} \times \mathbf{E}') - \mathbf{E}_0^* \cdot ({\nabla} \times \mathbf{H}')=\nabla \cdot (\mathbf{E}_0^*\times \mathbf{H}'-\mathbf{H}_0^*\times \mathbf{E}')+\mathbf{E}'\cdot ({\nabla} \times \mathbf{H}_0^*)-\mathbf{H}'\cdot ({\nabla} \times \mathbf{E}_0^*)
    \label{eq:div_formula}
\end{equation}
Using the identity in eq. (\ref{eq:div_formula}) and $\mathrm{\omega = \omega_0 + \Delta \omega}$ (where $\omega_0$ is the unperturbed complex resonance frequency and $\Delta\omega$ is the change in resonance frequency due to perturbation) to simplify eq. (\ref{eq:curl_add}) yields the following.
\begin{equation}
    i\Delta \omega(\mathbf{D}'\cdot\mathbf{E}_0^*+\mathbf{B}'\cdot\mathbf{H}_0^*)=i\omega_0[(\mathbf{D}_0^*\cdot\mathbf{E}'-\mathbf{D}'\cdot\mathbf{E}_0^*)-(\mathbf{B}'\cdot\mathbf{H}_0^*-\mathbf{B}_0^*\cdot\mathbf{H}')]+\nabla \cdot (\mathbf{E}_0^*\times \mathbf{H}'-\mathbf{H}_0^*\times \mathbf{E}')
    \label{eq:delta_omega}
\end{equation}
Integrating over the entire unit cell, we obtain the following.
\begin{equation}
\begin{aligned}
\int_{\mathrm{V}}\Delta \omega(\mathbf{D}'\cdot\mathbf{E}_0^*+\mathbf{B}'\cdot\mathbf{H}_0^*)\mathrm{dV}=\int_{\mathrm{V}}\omega_0[(\mathbf{D}_0^*\cdot\mathbf{E}'-\mathbf{D}'\cdot\mathbf{E}_0^*)-(\mathbf{B}'\cdot\mathbf{H}_0^*-\mathbf{B}_0^*\cdot\mathbf{H}')]\mathrm{dV}\\
-i\int_{\mathrm{V}}\nabla \cdot (\mathbf{E}_0^*\times \mathbf{H}'-\mathbf{H}_0^*\times \mathbf{E}')\mathrm{dV}
\label{Main_eq1}
\end{aligned}
\end{equation}
Using Gauss's theorem, the integral with the divergence term (second term on the RHS) in eq. \ref{Main_eq1} can be rewritten as a surface integral evaluated on the surface S of the unit cells of the metasurface,
\begin{equation}
    -i\int_{\mathrm{V}}\nabla \cdot [\mathbf{E}_0^*\times \mathbf{H}'-\mathbf{H}_0^*\times \mathbf{E}']\mathrm{dV}=-i\oint_{\mathrm{S}}\mathrm{\hat{n}}\cdot[\mathbf{E}_0^*\times \mathbf{H}'+\mathbf{E}'\times \mathbf{H}_0^*]\mathrm{dS}
    \label{eq:S6}
\end{equation}
where $\hat{\mathrm{n}}$ is the unit normal to the surface S. Furthermore, we assume that the perturbation in the hot spot at the center of the meta-atom has negligible impact on the fields at the unit cell walls, so that $\mathbf{E}'|_{\mathrm{walls}}\approx \mathbf{E}_0|_{\mathrm{walls}}$ and $\mathbf{H}'|_{\mathrm{walls}}\approx \mathbf{H}_0|_{\mathrm{walls}}$. Under this assumption, the RHS of eq. \ref{eq:S6} may be written as
\begin{equation}
    -i\oint_{\mathrm{S}}\mathrm{\hat{n}}\cdot[\mathbf{E}_0^*\times \mathbf{H}'+\mathbf{E}'\times \mathbf{H}_0^*]\mathrm{dS}\approx-2i\oint_{\mathrm{S}} \mathrm{\hat{n}}\cdot\mathrm{Re}(\mathbf{E}_0\times \mathbf{H}_0^*)\mathrm{dS}=-4i\oint_{\mathrm{S}} (\mathrm{\hat{n}}\cdot \mathbf{S}_0)\mathrm{dS}
    \label{Surface_int}
\end{equation}
where $\mathbf{S}_0$ is the Poynting vector and thus is purely real. Therefore, the surface integral in eq.(\ref{Surface_int}) calculates the change in the imaginary part of the resonance frequency ($\mathrm{Im}(\Delta \omega)$), which is also related to the quality factor of the metasurface resonance. However, since we are only interested in calculating the change in the real part of the resonance frequency ($\mathrm{Re}(\Delta \omega)$), we neglect the surface-integral term to obtain

\begin{equation}
    \frac{\mathrm{Re}(\Delta \omega)}{\mathrm{Re}(\omega_0)} = \frac{ \int_\mathrm{V} [(\mathbf{D}_0^*\cdot\mathbf{E}'-\mathbf{D}'\cdot\mathbf{E}_0^*)-(\mathbf{B}'\cdot\mathbf{H}_0^*-\mathbf{B}_0^*\cdot\mathbf{H}')]\mathrm{dV}}{\int_\mathrm{V} (\mathbf{D}'\cdot\mathbf{E}_0^*+\mathbf{B}'\cdot\mathbf{H}_0^*)\ \mathrm{dV}}
    \label{eq:delta_omega_frac_1}
\end{equation}
where $\omega_0$ is the resonance frequency of the unperturbed metasurface (which is complex for typically leaky metasurface resonances with the imaginary part related to the decay of the resonant fields), and $\mathrm{V}$ is the volume of the metasurface unit cell. Eq.(\ref{eq:delta_omega_frac_1}) accurately predicts the shifts in the resonance frequency for any perturbation to the meta-atom. However, it is of limited use since it requires knowledge of the complete fields in the perturbed structure, which are generally unknown.
\begin{figure}
\centering
    \includegraphics[height=48mm]{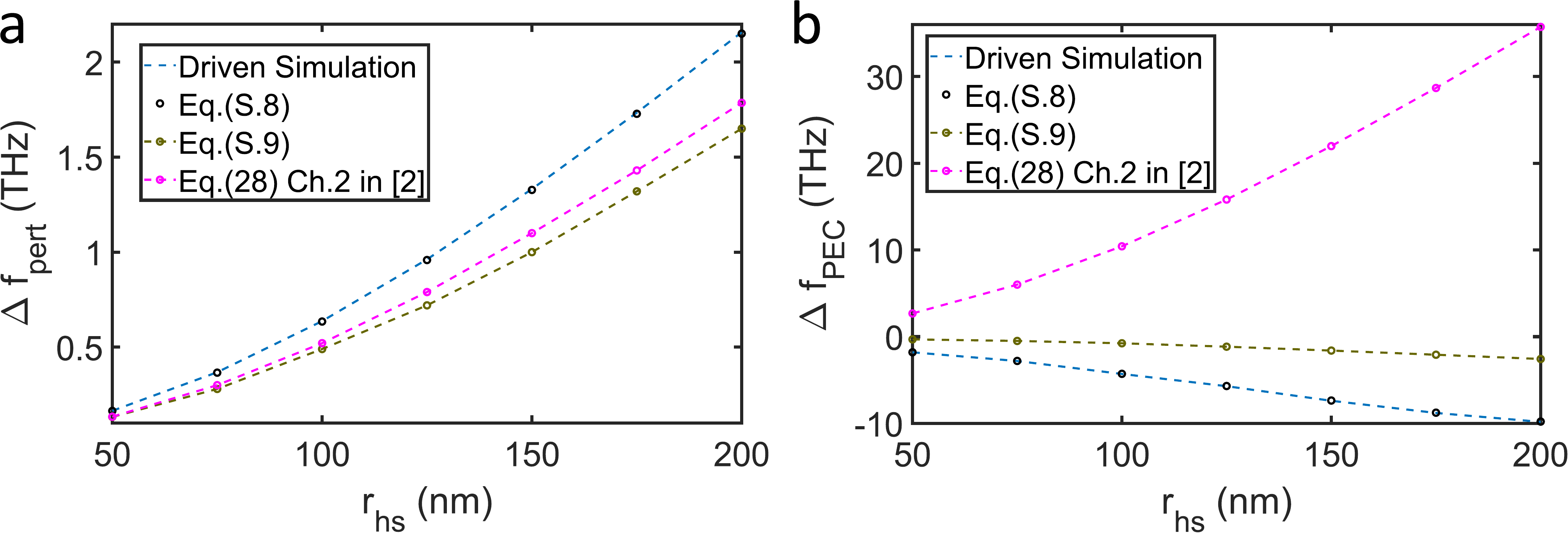}
\caption{Comparison of the resonance frequency shift predictions for $\bf{a.}$ $\mathrm{N_e=N_e^{pert}}$ = $\mathrm{1.6\times10^{19}cm^{-3}}$, and $\bf{b.}$ $\mathrm{N_e=N_e^{PEC}}$ = $\mathrm{3.2\times10^{20}cm^{-3}}$ for increasing hot spot radii.}
 \label{fig:1}
\end{figure}
To simplify Eq.(\ref{eq:delta_omega_frac_1}) to the lowest order in the perturbed fields, we assume that the fields in the perturbed volume (hot spot) is represented by a small cylinder of radius $r_{\rm hs}$, as shown in Fig.~1c of the main text, which can be broken up into infinitesimally thin annular shells with radii $0<\rho<\mathrm{r_{hs}}$, thicknesses $\mathrm{d}\rho$, and area $S_{\rho}$. Equation~(\ref{eq:delta_omega_frac_1}) may be further simplified by decomposing the fields into normal ($\perp$) and tangential ($\parallel$) components with respect to the surface $\mathrm{'S'}$ of the infinitesimally thin annular cylinder. Since $\mathbf{D}_{\perp}$, $\mathbf{E}_{\parallel}$, $\mathbf{H}_{\perp}$, and $\mathbf{B}_{\parallel}$ are continuous across an interface, all the fields in eq. (\ref{eq:delta_omega_frac_1}) are expressed in terms of these continuous components: $\mathbf{D}_{0} = \epsilon_0\varepsilon_\mathrm{ini} \mathbf{E}_{0\parallel}+\mathbf{D}_{0\perp}$, and $\mathbf{E}' = \mathbf{E}'_{\parallel} + ({1}/{\epsilon_0\varepsilon_\mathrm{fin}})\mathbf{D}'_{\perp}$. The second term inside the integral in the numerator of \ref{eq:delta_omega_frac_1} vanishes because the magnetic permeability is unchanged. Furthermore, since we consider small changes in permittivity and infinitesimally thin shells, we assume that $\mathbf{E}'_{\parallel}(\rho) =\mathbf{E}_{0\parallel}(\rho)$ and $\mathbf{D}'_{\perp}(\rho) = \mathbf{D}_{0\perp}(\rho)$ (azimuthal angle $\phi$ dependence and the variation of fields in the z direction have been implicitly assumed). Keeping only the unperturbed fields in the denominator of Eq.~\ref{eq:delta_omega_frac_1}, the following simplified expression for the frequency shift is obtained:
\begin{equation}
    \frac{\mathrm{Re}(\Delta \mathrm{f})}{\mathrm{Re}(\mathrm{f}_0)}=\frac{\int_0^\mathrm{r_{hs}} \left\{ \oint_{\mathrm{S}}\epsilon_0\left( [\mathrm{\varepsilon_{ini}}-\mathrm{\varepsilon_{fin}}] |\mathbf{E}_{0\parallel}(\rho)|^2
 -\left[\frac{1}{\mathrm{\varepsilon_{ini}}}-\frac{1}{\mathrm{\varepsilon_{fin}}} \right]|\mathbf{D}_{0\perp}(\rho)|^2\right)d\mathrm{S}\right\} \,\mathrm{d\rho}}{\int_\mathrm{V} (\mathbf{D}_0\cdot\mathbf{E}_0^*+\mathbf{B}_0\cdot\mathbf{H}_0^*)\ \mathrm{dV}}
    \label{eq:unpert_f}
\end{equation}
where f is the frequency in THz. The resonance frequency shifts predicted by eq.~(\ref{eq:delta_omega_frac_1}), and eq.~(\ref{eq:unpert_f}) are then compared to driven simulations in COMSOL Multiphysics and also with the resonance frequency shifts predicted using eq.~(28) in chapter 2 of ref.~ \cite{Joannopoulos1995} for reference. Fig.~\ref{fig:1}a, and \ref{fig:1}b show the calculated shift in the metasurface resonance frequency using these various formulae for different hot spot radii ($\mathrm{r_{hs}}$) when $\mathrm{N_e=N_e^{pert}}$ = $\mathrm{1.6\times10^{19}cm^{-3}}$, and $\mathrm{N_e=N_e^{PEC}}$ = $\mathrm{3.2\times10^{20}cm^{-3}}$, respectively. The shifts predicted by eq.~(\ref{eq:delta_omega_frac_1}), using knowledge of the fields in the perturbed system, are the most accurate and in accordance with the COMSOL simulations. From Fig.~\ref{fig:1}a, we observe that the resonance frequency shifts predicted from eq.(\ref{eq:unpert_f}) using only the unperturbed fields are reasonably accurate and are similar to the shifts predicted using eq.~(28) in chapter 2 of Ref.~ \cite{Joannopoulos1995} for the case of small perturbation (i.e., Fig.~\ref{fig:1}a, where $\mathrm{N_e=N_e^{pert}}$). For large perturbations, when the hot spot permittivity becomes negative (i.e., Fig.~\ref{fig:1}b, where $\mathrm{N_e=N_e^{PEC}}$), eq.~(\ref{eq:delta_omega_frac_1}) is still the most accurate; however, in this case, eq.~(\ref{eq:unpert_f}) predicts shifts on the correct side of the unperturbed frequency while eq.~(28) in chapter 2 of Ref.~ \cite{Joannopoulos1995} does not. This shows that eq.~(\ref{eq:unpert_f}) can accurately predict the nature of the resonance frequency shifts for all perturbations.

\section{Manipulating Magnetic Energy Density}

\begin{figure}[t]
\centering
    \includegraphics[width=120mm]{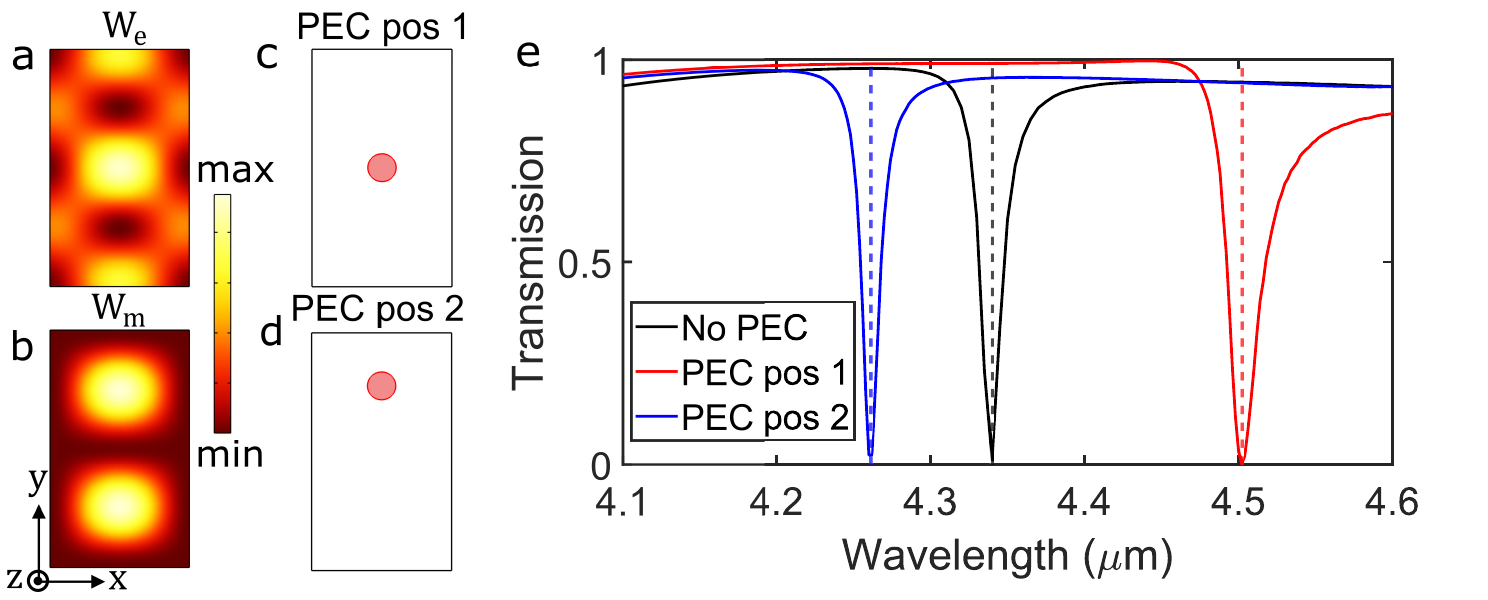}
    \caption{$\bf{a.}$ ($\bf{b.}$) Normalized electric (magnetic) energy density of the ED resonance for the unperturbed metasurface. $\bf{c.}$ Schematic showing an effectively metallized (PEC) but smaller hot spot ($\mathrm{r_{hs}=100 nm}$) placed in the center of the Ge block, where $\mathrm{W_e}$ is dominant (position 1). $\bf{d.}$ Same as $\bf{c}$ but with the hot spot now placed in a location, where $\mathrm{W_m}$ is dominant (position 2). $\bf{e.}$ Metasurface transmission spectrum for the cases where the metasurface is unperturbed (black), PEC is in position 1 (red), and PEC is in position 2 (blue).}
    \label{fig:2}
\end{figure}

The electromagnetic energy density of a resonant mode is unevenly distributed within the meta-atom, with the electric and magnetic energy densities dominant in different spatial regions. Therefore, despite considering a metasurface comprising nonmagnetic material ($\mu_{\mathrm{r}}=1$), it is possible to modify the magnetic energy density of the resonant mode, e.g., by effectively metallizing or introducing PEC in a specific region where the magnetic energy density dominates the electric energy density. The resulting frequency shift $\Delta \omega_{\rm PEC}$ of the metasurface resonance is given by
\begin{equation}
\begin{gathered}
    \frac{\mathrm{Re}(\Delta\omega^2_{\rm PEC})}{\mathrm{Re}(\omega^2_0)} =  \int_{\Delta \mathrm{V}} \left( \mu_0\left| \mathbf{H}_0 \right|^2 - \epsilon_0\varepsilon_\infty\left| \mathbf{E}_0 \right|^2 \right) \mathrm{dV}\\ \int_\mathrm{V} \mathbf{H}_0^*.\mathbf{H}_0\mathrm{ dV}=\frac{1}{\mu_0},\ \int_\mathrm{V} \mathbf{E}_0^*.\mathbf{E}_0\mathrm{ dV}=\frac{1}{\epsilon_0\varepsilon_\infty}
    \end{gathered}
    \label{eq:Slater}
\end{equation}
where $\mathbf{E}_0$ and $\mathbf{H}_0$ are the normalized unperturbed electric and magnetic fields, respectively, and $\Delta\mathrm{V}$ is the PEC volume inside each meta-atom. Assuming a dispersion-free high-frequency relative permittivity for Ge (i.e.,~$\mathrm{d}\varepsilon_\infty/{\mathrm{d}\omega}=0$) and absence of free carriers ($\omega_\mathrm{p}=0$), the above choice of normalization gives the magnetic and electric energy density as $\mathrm{W_m}=\mu_0|\mathbf{H}_0|^2/2$ and $\mathrm{W_e}=\epsilon_0\varepsilon_\infty|\mathbf{E}_0|^2/2$, respectively. According to Eq.~(\ref{eq:Slater}) in the main text, the introduction of a PEC in a region of a resonator where the magnetic (electric) energy density is dominant would result in a blue shift (red shift) of the resonance \cite{Slater1946, Ma2017PRB}.

Thus, when a strongly metallized (PEC) but smaller hot spot of radius $\mathrm{r_{hs}}=$100 nm is introduced in the center of the Ge block (Fig.~\ref{fig:2}c), it occupies a region of high electric energy density. This expels out the electric field from the PEC volume and increases the effective capacitance while redshifting the ED resonance from 4.34 $\mu \mathrm{m}$ to 4.5 $\mu \mathrm{m}$ as shown in Fig.~\ref{fig:2}e. We have seen this effect in the previous sections, but note that our choice of a smaller hot spot region reduces the extent of the redshift of the ED resonance. However, when a PEC is introduced in an off-center region of the Ge block where the magnetic energy density is dominant (Fig.~\ref{fig:2}d), the resonance blueshifts to 4.26 $\mu$m (Fig.~\ref{fig:2}e). Since the PEC now expels the magnetic field, it reduces the effective inductance of the meta-atom due to excessive FCs that shield the magnetic flux, resulting in a blueshift of the ED resonance.

\section{Meta-atom Homogenization: Effective Permittivity Model}

\begin{figure}
\centering
    \includegraphics[height=90mm]{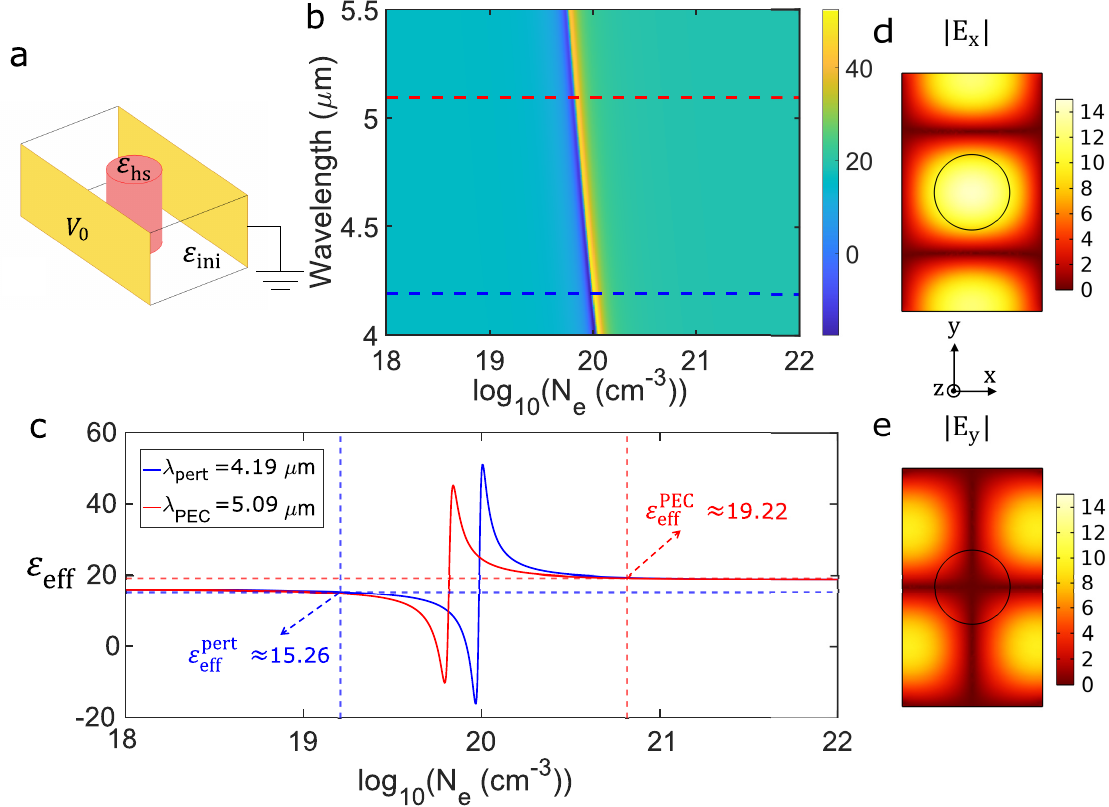}
 \caption{$\bf{a.}$ Geometry for the effective capacitor model. The parallel plates are shown as golden surfaces, and the hot spot at the center is red. $\bf{b.}$ The effective permittivity of the Ge block as a function of local hot spot carrier density and wavelength. The black, blue, and red dashed lines in $\bf{c.}$ represent the $\mathrm{\lambda_i}$, $\mathrm{\lambda_{pert}}$, and $\mathrm{\lambda_{PEC}}$ wavelengths, respectively. $\bf{c.}$ $\mathrm{\varepsilon_{eff}^{xx}}$ as a function of $\mathrm{N_e}$ for two different values of $\lambda=\lambda_{\mathrm{pert}}$ (blue) and $\lambda=\lambda_{\mathrm{PEC}}$ (red). Note that $\varepsilon_\mathrm{eff}^\mathrm{pert}<\varepsilon_\mathrm{ini}\approx16<\varepsilon_\mathrm{eff}^\mathrm{PEC}$. $\bf{d.}$ $|\mathrm{E_x}|$ field enhancement in the Ge block with the hot spot region marked using a bold circle. $\bf{e.}$ Same as $\bf{d}$ but for $|\mathrm{E_y}|$.}
 \label{fig:3}
\end{figure}

In this section, we provide further insight into the behavior of the metasurface resonance upon FC generation, particularly the redshifting of the ED resonance at high values of $\mathrm{N_e}>10^{20}\ \mathrm{cm^{-3}}$. We employ an electrostatic capacitor model \cite{Shvets2004PRL, URZHUMOV_Shvets_2008} to homogenize the Ge resonator and the hot spot and calculate its effective permittivity ($\mathrm{\varepsilon_{eff}}$). The Ge block and the hot spot are modeled as a parallel plate capacitor, and we calculate the total charge ($\mathrm{Q}$) accumulated on the two longer faces ($\parallel$ to the $y$-$z$ plane) when a constant potential difference $\mathrm{V_0}$ is applied across them. The capacitor plates have an area $\mathrm{A=w_y\times h}$ with separation $\mathrm{w_x}$ as shown in Fig.~\ref{fig:3}a. The longer sides of the meta-atom are chosen as the parallel plates of the capacitor in our model because changes in the permittivity of the hot spot located in the center of the Ge block affect $\mathrm{E_x}$ more than $\mathrm{E_y}$. This can be ascribed to the fact that the maximum (minimum) of $|\mathrm{E_x}|\ (|\mathrm{E_y}|)$ coincides with the location of the hot spot, as shown in Fig.~\ref{fig:3}d(e). Then, after calculating $\mathrm{Q}$, we obtain $\mathrm{\varepsilon_{eff}^{\mathrm{xx}}}$ for the x-polarized incidence
\begin{equation}
    \mathrm{\varepsilon_{eff}^{\mathrm{xx}} = \frac{Q w_x}{\epsilon_0 w_y h V_0}}.
\end{equation}

Fig.~\ref{fig:3}b shows the value of $\mathrm{\varepsilon_{eff}^{\mathrm{xx}}}$ calculated using the capacitor model for the range of $\mathrm{N_e}$ and $\lambda$ studied in Fig.~2a of the main text. Fig.~\ref{fig:3}c shows the line cuts of Fig.~\ref{fig:3}b where we plot the variation in $\mathrm{\varepsilon_{eff}}$ as a function of $\mathrm{N_e}$ for three different wavelengths $\mathrm{\lambda_i}$ (black), $\mathrm{\lambda_{pert}}$ (blue) and $\mathrm{\lambda_{PEC}}$ (red). For all the values of $\lambda$, we find that at low values of $\mathrm{N_e}$, $\mathrm{\varepsilon_{eff}^{xx}=\varepsilon_{eff}^{i}=\varepsilon_\infty}$. $\mathrm{\varepsilon_{eff}^{xx}}$ rapidly decreases from $\varepsilon_\infty$ to 0 for $10^{19}\ \mathrm{cm^{-3}}\leq\mathrm{N_e} \leq10^{20}\ \mathrm{cm^{-3}}$. For values of $\mathrm{N_e}\geq 10^{20}\ \mathrm{cm^{-3}}$, while the local permittivity of the hot spot becomes negative, the effective permittivity of the homogenized medium $\mathrm{\varepsilon_{eff}^{xx}}$ increases rapidly and saturates at approximately $\mathrm{\varepsilon_{eff}^{xx}}\approx 19$. The changes in $\mathrm{\varepsilon_{eff}^{\mathrm{xx}}}$ as $\mathrm{N_e}$ crosses $\sim10^{20}\ \mathrm{cm^{-3}}$ account for changes in the metasurface resonance wavelength seen in Fig.~2a of the main text as $\mathrm{N_e}$ crosses $10^{20}\ \mathrm{cm}^{-3}$ since the resonance wavelength $\lambda_r\propto\sqrt{\mathrm{C_{eff}}}\propto\sqrt{\mathrm{\varepsilon_{eff}^{\mathrm{xx}}}}$ where $\mathrm{C_{eff}}$ is the effective capacitance. The intersection of the dashed blue (red) lines shows the value of the effective permittivity for $\lambda=\lambda_{\mathrm{pert}} \ (\lambda_{\mathrm{PEC}})$ and $\mathrm{N_e=N_e^{pert}\ (N_e^{PEC})}$. Thus, localized FC generation in the hot spots initially decreases the effective permittivity of the meta-atom and blue-shifts the metasurface resonance. Further increasing the FC density creates a localized region of high negative permittivity ($\mathrm{\varepsilon_{hs}}$), thereby increasing the effective permittivity and red-shifting the metasurface resonance. Since the effective capacitor model is electrostatic, it can only be used to determine the effective permittivity for modifications that primarily affect the electric energy.

\section{Keldysh Photoionization Model}

We used the Keldysh model \cite{Keldysh1965ojf} to determine the time evolution of the FC density ($\mathrm{N_e}(\mathbf{r},\mathrm{t)}$) inside the hot spot as a consequence of illumination by a high-intensity, ultra-short pump pulse. We modeled the laser-induced ionization of Ge by calculating the FC density due to the photoionization process ($\mathrm{R_{PI}}$),

\begin{equation}
    \frac{\partial \mathrm{N_e}(\mathbf{r},\mathrm{t})}{\partial \mathrm{t}} = \mathrm{R_{PI}(I[t])}-\frac{\mathrm{N_e}(\mathbf{r},\mathrm{t})}{\tau_\mathrm{e-h}}
    \label{Keldysh}
\end{equation}

where I[t] is the time-dependent intensity of the pump pulse and $\tau_\mathrm{e-h}$ is the electron-hole recombination time. Here, we consider photoionization as the primary mechanism for free-carrier generation in Ge; other processes, such as impact/avalanche ionization by the photoionized electrons, will be included in our future calculations. The pump pulse parameters and the germanium optical properties are shown in tabular form below \cite{van2011book, Amotchkina2020, Ge_dynamics2017}.

\begin{center} 
\centering
 \begin{tabular}{ | m{12em} | m{2cm}| m{5.5cm} | } 
 \hline
 \bf{Parameter} & \bf{Symbol} & \bf{Value} \\ 
 \hline\hline
 Pump wavelength & $\mathrm{\lambda_p}$ & 1580 nm \\ 
 \hline
 Pulse width & $\mathrm{\tau_{p}}$ & 8 fs \\ 
 \hline
 Band Gap of Ge & $\Delta$ & 0.8 eV \\ 
 \hline
  Reduced effective mass & $\mathrm{m_e}$ & 0.041 $\mathrm{m_0, \hspace{0.2cm} m_0=9.11\times 10^{-31}\hspace{0.1cm}kg}$ \\ 
 \hline
  Refractive index (at $\mathrm{\lambda}$) & $\mathrm{n_0}$ & 4.2 \\ 
 \hline
 Pump Irradiance & I[t] & $\mathrm{I[t]=2I_{avg}e^{-(t-\tau_2)^2/\tau_{p}^2}},\ \tau_2=3\ \mathrm{ps}$ \\ 
 \hline
 e-h Recombination time & $\tau_\mathrm{e-h}$ & 100 ps \\
 \hline
\end{tabular}
\end{center}

\begin{figure}[t]
 \centering
 
    \includegraphics[width=0.6\linewidth]{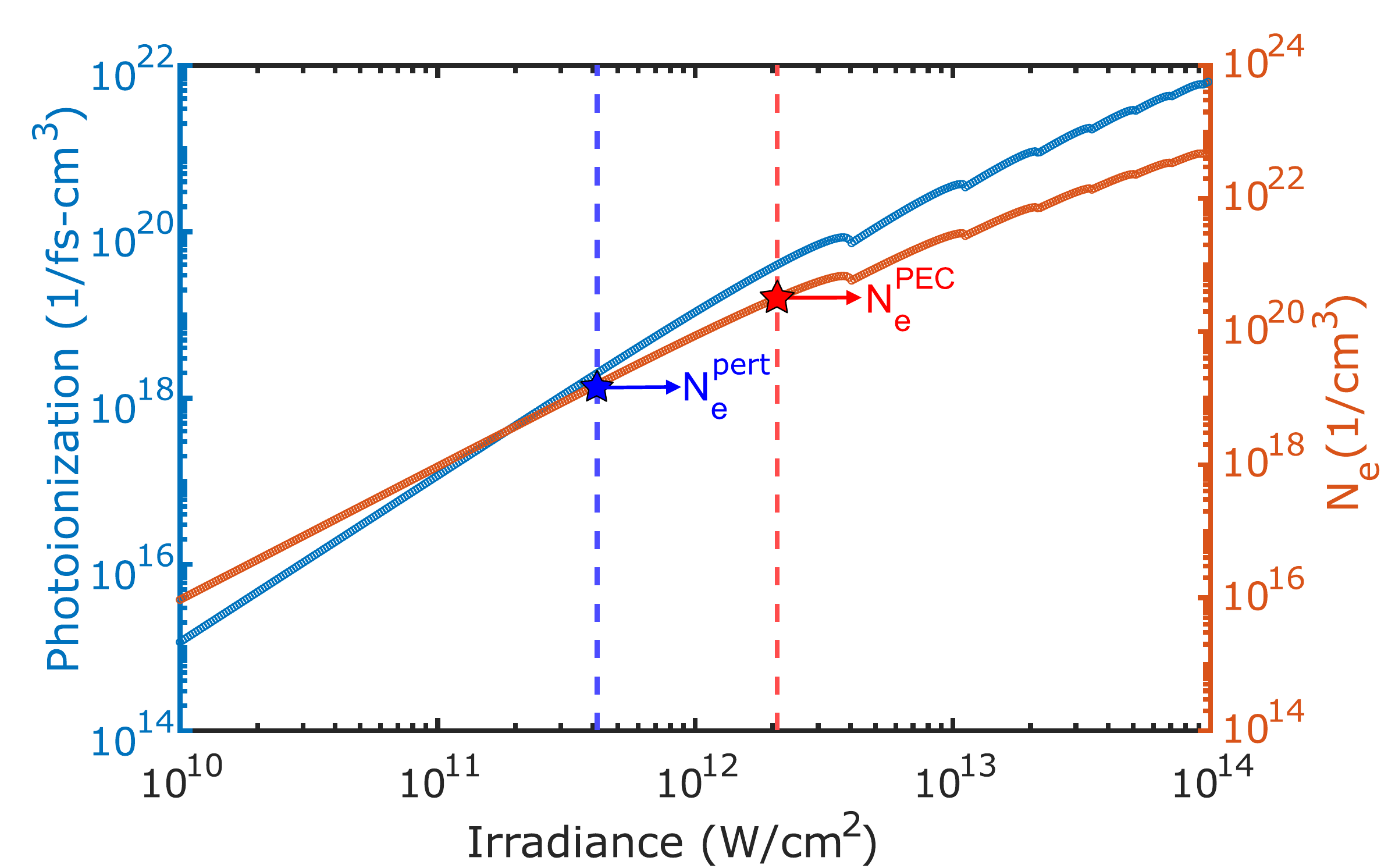}
    
 \caption{Photoionization (blue) rate for Ge, calculated using the Keldysh model for the given pump pulse parameters. Carrier concentration (red, $\mathrm{N_e}$) as a function of input irradiance; the red (blue) horizontal dashed line represents the average irradiance 2.075 $\mathrm{TW/cm^{2}}$ (416 $\mathrm{GW/cm^2}$) required for the red-shifting (blue-shifting) of resonance wavelength to $\mathrm{\lambda_{PEC}}$ ($\mathrm{\lambda_{pert}}$).}
 \label{fig:4}
\end{figure}

The photoionization rate calculated using the Keldysh model is given by \cite{Gruzdev2014SPIE}
\begin{equation}
    \mathrm{R_{PI} = 2.\frac{2\omega}{9\pi} \left(\frac{\sqrt{1+\gamma^2}}{\gamma}.\frac{m_e\omega}{\hbar}\right).Q_k\left(\gamma,\frac{\Delta_{NP}}{\hbar \omega} \right).exp\left[-\pi\left<\frac{\Delta_{NP}}{\hbar \omega}+1 \right>.\frac{K(\phi)-E(\phi)}{E(\theta)} \right]}
\end{equation}
where a factor of 2 is introduced to account for the electron spin degeneracy, and $\mathrm{Q_k}$ represents a slow-varying amplitude function and is written as
\begin{equation}
    \mathrm{Q_k(\gamma,x) = \sqrt{\frac{\pi}{2.K(\theta)}}.\sum_{n=0}^{+\infty} exp\left[-\pi.\frac{K(\phi)-E(\phi)}{E(\theta)}.n \right].\Phi \left[\sqrt{\frac{\pi^2(<x+1>-x+n)}{2.K(\theta).E(\theta)}} \right],}
\end{equation}
where, $\mathrm{K(x)}$ and $\mathrm{E(x)}$ are complete elliptic integrals. $\mathrm{\theta}$, $\mathrm{\phi}$, and $\mathrm{\gamma}$ are given as
\begin{equation}
    \mathrm{\theta=\frac{1}{1+\gamma^2},\hspace{1cm} \phi=\frac{\gamma}{1+\gamma^2}, \hspace{1cm} \gamma=\frac{\omega\sqrt{m\Delta}}{eF}}
\end{equation}
$\mathrm{\Phi(x)}$ and $\mathrm{\Delta_{NP}}$ represent the Dawson integral and the laser-modified band gap, which are given as
\begin{equation}
    \mathrm{\Phi(x) = \int_{0}^{x} (\chi^2-x^2) \,d\chi} ;\ \ \mathrm{\Delta_{NP}=\frac{2}{\pi}\Delta.\left[\frac{\sqrt{1+\gamma^2}}{\gamma}.E\left(\frac{1}{\sqrt{1+\gamma^2}} \right) \right]},
\end{equation}
where $\mathrm{\omega=2\pi c/\lambda_p}$ is the angular frequency of the pump pulse, $\mathrm{e}$ is the electronic charge, $\mathrm{F}$ is the local electric field amplitude, and $\mathrm{<x>}$ the integer part of x. The photoionization rate in bulk Ge, calculated using the Keldysh model for different pump irradiances, is shown in Fig.~\ref{fig:4}. Using a source-driven simulation, we observed that the Ge blocks do not exhibit collective behavior at the pump frequency. Therefore, the photoionization rate for the Ge blocks is assumed to be the same as the rates for bulk germanium. The free-electron concentration in the hot spot for various pump intensities is also shown in Fig.~\ref{fig:4}. 

\section{Simulation Setup and FC Generation}

\begin{figure}[t]
 \centering
 
    \includegraphics[width=0.65\linewidth]{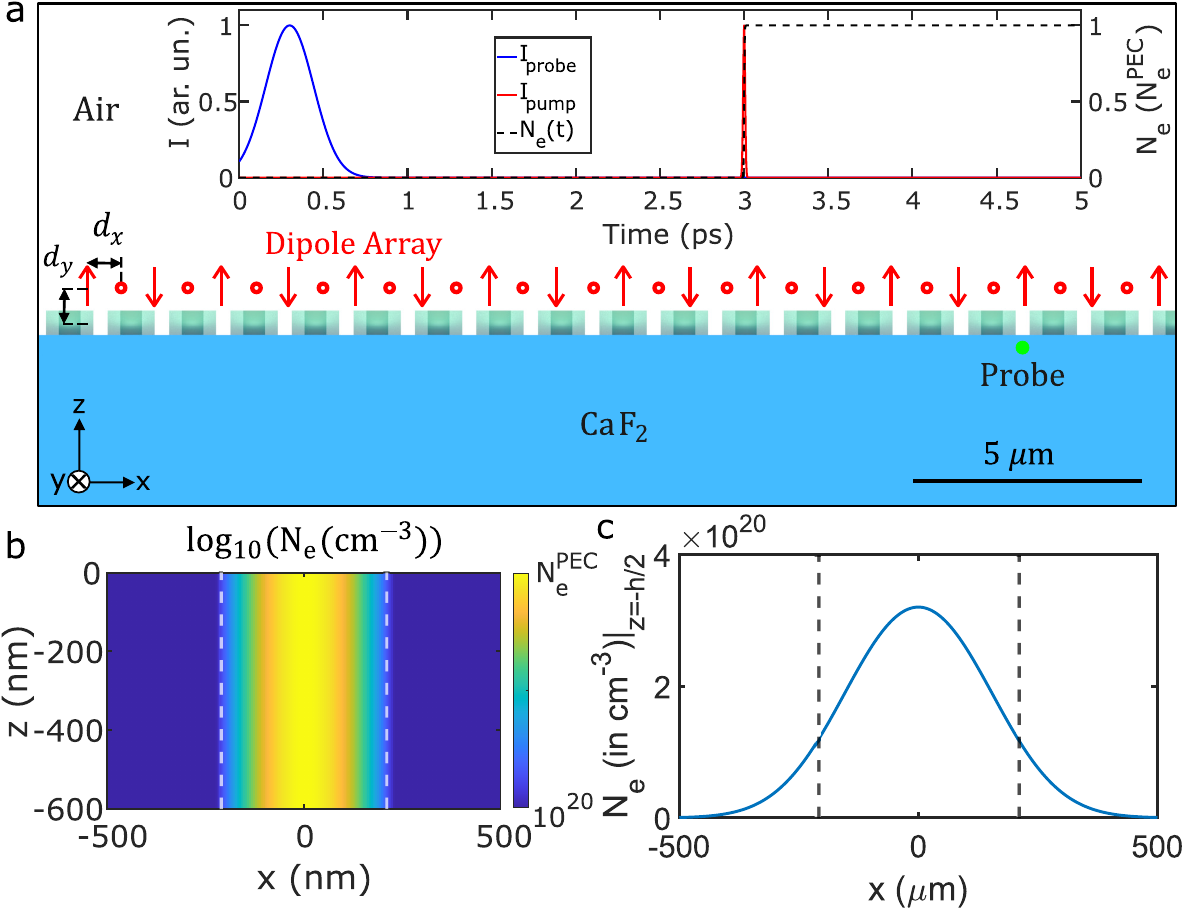}
    
 \caption{$\bf{a.}$ Setup used for TI simulations where the shaded red region shows the effective cylindrical hot spots of radius 210 nm inside each meta-atom. Probe points are located in the substrate (green circles). MGW is launched using a dipole array (red arrows) separated by $d_x=\pi/2\mathrm{k_x}$ and at a height $d_y=\lambda_0/4$ above the metasurface; the length of the arrows represents the magnitude of the dipole moment. Inset: Pump pulse (red) and dipolar excitation (blue) intensity and the temporal variation of hot spot FC density (black dashed). $\bf{b.}$ Spatial distribution of the FC density in the meta-atom during the photoionization process created using an average irradiance of $\mathrm{I_{avg}^{PEC}}=2.075\ \mathrm{TW/cm^2}$; the dashed vertical lines enclose the effective hot spot region. $\bf{c.}$ FC density in a horizontal cut plane of the 2D plot in $\bf{a}$ at z=-300 nm.}
 \label{fig:5}
\end{figure}

 The simulation setup used to implement a TI during MGW propagation is shown in Fig.~\ref{fig:5}a, where the red volume in each meta-atom denotes the hot spot region. The probe points at which we inspect the signal are marked with yellow circles in Fig.~\ref{fig:5}a. The red arrows indicate a spatial array of phased dipoles with spacing $d_x=\pi/2\mathrm{k_x}$ located at a height $d_y=\lambda_0/4$ above the metasurface with dipole moments pointing in the $\mathrm{\hat{z}}$ direction. The normalized pump/probe intensities and the temporal evolution of $\mathrm{N_e}$ in the hot spot are shown in the inset of Fig.~\ref{fig:5}a. 
 
 Furthermore, we estimate the spatial inhomogeneity of the FC generation process in the hot spot and use it to accurately model the TI. For the pump pulse parameters used, the Keldysh parameter is less than 1 ($\gamma\ll1$), indicating that we are in the strong tunneling ionization regime (a highly nonlinear process). Therefore, even when the radius of the focused laser spot on each meta-atom is taken to be diffraction-limited ($\sim \lambda_\mathrm{p}/2$), the spatial distribution of the free-carriers created does not follow the spatial distribution of the pump intensity in the meta-atom. This implies that the FCs are more strongly localized in the meta-atom due to the nonlinear dependence of the tunneling ionization rate on the pump intensity, which is higher at the center of the focused spot. 

To obtain the spatial distribution of the FCs, we use the Keldysh model to calculate the photoionized FCs created by the pump pulse, whose intensity varies spatially within the meta-atom. On the surface of each meta-atom (i.e., in the $x$-$y$ plane), we assume a Gaussian distribution of the electric field due to the pump pulse with the beam waist equivalent to $w_0\sim\lambda_\mathrm{p}/4=395\ \mathrm{nm}$ (diffraction-limited focused spot)
\begin{equation}
    \mathrm{E}(x,y,z)\propto\exp{(-(x^2+y^2)/w_0^2)}\Rightarrow\mathrm{I}(x,y,z)=\mathrm{I_{avg}^{PEC}}\exp{(-2(x^2+y^2)/w_0^2)}.
    \label{eq:sp_I}
\end{equation}
where $\mathrm{I_{avg}^{PEC}}$ is the average pump intensity required to create a FC density of $\mathrm{N_e^{PEC}}$. We used the spatially varying pump intensity from eq.~(\ref{eq:sp_I}) in the Keldysh model to calculate the spatial dependence of the photoionized electrons and fit them to a Gaussian distribution (see Fig.~\ref{fig:5}b) 
\begin{equation}
    \mathrm{N_e}(x,y,z)=\mathrm{N_{e}^{PEC}}\exp{(-(x^2+y^2)/\sigma^2)}
    \label{eq: N_e_fit}
\end{equation}
Fitting the FC distribution obtained from the Keldysh model to eq.~(\ref{eq: N_e_fit}), the Gaussian waist of the distribution, $\sigma\sim210\ \mathrm{nm}\ll w_0$, is taken as the effective radius of the hot spot in all the frequency and time domain simulations (see Fig.~\ref{fig:5}c). However, we used the complete spatial FC density profile while calculating the effects of losses and spatially inhomogeneous FC generation on the rectified magnetic field in Section 4.3.2 of the main text.

\section{Energy Relations in a Time-Varying Drude-Lorentz Medium}

In a time-varying Drude-Lorentz medium, the total electron density $\mathbf{J}_\mathrm{e}$ can be written as the sum of pre-existing free electron density $\mathbf{J}^{(1)}_\mathrm{e}$ and newly created free electron density $\mathbf{J}^{(2)}_\mathrm{e}$ (see Section 3.2 in the main text).
\begin{equation}
\begin{gathered}
    \mathbf{J}_\mathrm{e}(\mathrm{t})=\frac{\partial \mathbf{P}_\mathrm{e}}{\partial \mathrm{t}}= \mathbf{J}^\mathrm{(1)}_\mathrm{e}(\mathrm{t}) + \mathbf{J}^\mathrm{(2)}_\mathrm{e}(\mathrm{t}) =-\mathrm{eN^{(1)}_e \mathbf{v}^{(1)}_e(t)}-\mathrm{e}\int_{-\infty}^\mathrm{t}\frac{\partial\mathrm{N^{(2)}_e}}{\partial\mathrm{t'}}\mathbf{v}^{(2)}_\mathrm{e}(\mathrm{t,t'})\mathrm{dt'}
    \\
    \mathbf{J}^{(1)}_\mathrm{e}(\mathrm{t})=-\frac{\mathrm{e}^2}{\mathrm{m_e}}{\mathrm{N^{(1)}_e}}\mathbf{A}(\mathrm{t});\ \ 
    \mathbf{J}^{(2)}_\mathrm{e}(\mathrm{t})=-\frac{\mathrm{e}^2}{\mathrm{m_e}}{\mathrm{N^{(2)}_e}}\mathbf{A}(\mathrm{t})+\frac{\mathrm{e}^2}{\mathrm{m_e}}\int_{-\infty}^\mathrm{t}\frac{\partial\mathrm{N^{(2)}_e}}{\partial \mathrm{t'}}\mathbf{A}(\mathrm{t'})\mathrm{dt'}
    \end{gathered}
    \label{eq:J_e}
\end{equation}
Now, we use Poynting's theorem in conjunction with Maxwell's equations to write the time derivative of the total energy density $'\mathrm{U}'$ in a dispersive medium as \cite{landau2003electrodynamics}
\begin{equation}
\begin{gathered}
    \mathrm{\frac{\partial U}{\partial t}=\frac{\partial U_{Field}}{\partial t}+\frac{\partial U_{Carrier}}{\partial t}}=-\nabla \cdot(\mathbf{E}\times\mathbf{H})=\mathbf{E}\cdot(\nabla\times\mathbf{H})-\mathbf{H}\cdot(\nabla\times\mathbf{E})
    \\ 
    \Rightarrow \mathrm{\frac{\partial U}{\partial t}}=\frac{\partial}{\partial \mathrm{t}}\left(\frac{1}{2}\epsilon_0\varepsilon_\infty|\mathbf{E}|^2 + \frac{1}{2}\mu_0 |\mathbf{H}|^2\right)+\mathbf{E}\cdot\frac{\mathrm{\partial}\mathbf{P}_\mathrm{e}}{\mathrm{\partial t}}
    \\
    \mathrm{U_{Field}}=\frac{1}{2}\epsilon_0\varepsilon_\infty|\mathbf{E}|^2 + \frac{1}{2}\mu_0 |\mathbf{H}|^2;\ \ \ \  \mathrm{U_{Carrier}}=\int_{-\infty}^\mathrm{t}\mathbf{E}\cdot\frac{\mathrm{\partial}\mathbf{P}_\mathrm{e}}{\mathrm{\partial t'}}\mathrm{dt'}
    \label{EM energy}
    \end{gathered}
\end{equation}
Using $\mathbf{E}=-\partial\mathbf{A}/\partial \mathrm{t}$, and from eq.~(\ref{eq:J_e}), the expression for ${\partial \mathbf{P}_\mathrm{e}}/{\partial \mathrm{t}}$ in the last term on the right side of eq.~(\ref{EM energy}), we get
\begin{equation}
    \mathbf{E}\cdot\frac{\mathrm{\partial}\mathbf{P}_\mathrm{e}}{\mathrm{\partial t}}=-\frac{\mathrm{\partial}\mathbf{A}}{\mathrm{\partial t}}\cdot\mathbf{J}_\mathrm{e}=\frac{\mathrm{e^2N^{(1)}_e}}{2\mathrm{m_e}}\frac{\mathrm{\partial}|\mathbf{A}|^2}{\mathrm{\partial t}}+\frac{\mathrm{e^2N^{(2)}_e(t)}}{2\mathrm{m_e}}\frac{\mathrm{\partial}|\mathbf{A}|^2}{\mathrm{\partial t}}-\frac{\mathrm{e^2}}{\mathrm{m_e}}\frac{\mathrm{\partial}\mathbf{A}}{\mathrm{\partial t}}\cdot\int_{-\infty}^\mathrm{t}\frac{\partial\mathrm{N^{(2)}_e}}{\partial\mathrm{t'}}\mathbf{A}(\mathrm{t'})\mathrm{dt'}
\end{equation}
Integrating the equation above and further simplifying using integration by parts yields
\begin{equation}
\begin{aligned}
    \mathrm{U_{Carrier}}=\int_{-\infty}^\mathrm{t}\mathbf{E}\cdot\frac{\mathrm{\partial}\mathbf{P}_\mathrm{e}}{\mathrm{\partial t'}}\mathrm{dt'}=&\left[\frac{\mathrm{e^2N^{(1)}_e |\mathbf{A}|^2}}{2\mathrm{m_e}}+\frac{\mathrm{e^2N^{(2)}_e(t) |\mathbf{A}|^2}}{2\mathrm{m_e}}-\frac{\mathrm{e^2}}{2\mathrm{m_e}}\int_{-\infty}^{\mathrm{t}}\frac{\partial\mathrm{N^{(2)}_e}}{\partial\mathrm{t'}}|\mathbf{A}(\mathrm{t'})|^2\mathrm{dt'}\right]\\ &-\left[\mathbf{A}(\mathrm{t})\cdot\int_{-\infty}^\mathrm{t}\frac{\partial\mathrm{N^{(2)}_e}}{\partial\mathrm{t'}}\mathbf{A}(\mathrm{t'})\mathrm{dt'}-\frac{\mathrm{e^2}}{\mathrm{m_e}}\int_{-\infty}^{\mathrm{t}}\frac{\partial\mathrm{N^{(2)}_e}}{\partial\mathrm{t'}}|\mathbf{A}(\mathrm{t'})|^2\mathrm{dt'}\right]
\end{aligned}
\end{equation}
Further simplifying the above expression, $\mathrm{U_{Carrier}}$ can be rewritten as
\begin{equation}
\begin{gathered}
    \mathrm{U_{Carrier}}=\frac{1}{2}\epsilon_0\omega_\mathrm{p}^2|\mathbf{A}|^2+\int_{-\infty}^{\mathrm{t}}\frac{\partial\mathrm{N^{(2)}_e}}{\partial\mathrm{t'}}\frac{|\mathrm{e}\mathbf{A}(\mathrm{t'})|^2}{2\mathrm{m_e}}\mathrm{dt'}-\mathbf{A}(\mathrm{t})\cdot\left(\frac{\mathrm{e^2}}{\mathrm{m_e}}\int_{-\infty}^\mathrm{t}\frac{\partial\mathrm{N^{(2)}_e}}{\partial\mathrm{t'}}\mathbf{A}(\mathrm{t'})\mathrm{dt'}\right)
    \label{eq:KE}
    \end{gathered}
\end{equation}
Integrating eq.(\ref{EM energy}) and using the expressions for $\mathrm{U_{Carrier}}$ from eq.~(\ref{eq:KE}), yields
\begin{equation}
\begin{aligned}
    \mathrm{U}=\frac{1}{2}\epsilon_0|\mathbf{E}|^2+\frac{1}{2}\mu_0 |\mathbf{H}|^2+\frac{1}{2}\epsilon_0(\varepsilon_\infty-&1)|\mathbf{E}|^2+\frac{1}{2}\epsilon_0\omega^2_\mathrm{p}|\mathbf{A}|^2\\ +\int_{-\infty}^{\mathrm{t}}\frac{\partial\mathrm{N^{(2)}_e}}{\partial\mathrm{t'}}\frac{|\mathrm{e}\mathbf{A}(\mathrm{t'})|^2}{2\mathrm{m_e}}\mathrm{dt'}-&\mathbf{A}(\mathrm{t})\cdot\left(\frac{\mathrm{e^2}}{\mathrm{m_e}}\int_{-\infty}^\mathrm{t}\frac{\partial\mathrm{N^{(2)}_e}}{\partial\mathrm{t'}}\mathbf{A}(\mathrm{t'})\mathrm{dt'}\right)
\end{aligned}
\label{Energy}
\end{equation}
We use eq.~(\ref{Energy}) to investigate the effects of a TI on total energy in a Drude-Lorentz medium varying in time, where the free electrons that create a TI are generated at rest. A detailed analysis of all the terms in eq.~(\ref{Energy}) is given in Sections 3.3 and 4.2 of the main text.

\section{Quasistatic Magnetic Field Amplitude}

We use Ampere's law (eq.~(5) in the main text) along with the expression for the current density from eq.~(\ref{eq:J_e}) to write
\begin{equation}
    \begin{aligned}
        \nabla(\nabla\cdot\mathbf{A})-\nabla^2\mathbf{A} +\frac{\varepsilon_\infty}{\mathrm{c^2}} \frac{\partial^2 \mathbf{A}}{\partial \mathrm{t}^2} + \frac{\omega_\mathrm{p}^2}{\mathrm{c}^2}\mathbf{A}-\frac{\mathrm{e}^2}{\epsilon_0\mathrm{m_e c^2}}\int_{-\infty}^\mathrm{t}\frac{\partial\mathrm{N^{(2)}_e}}{\partial \mathrm{t'}}\mathbf{A}(\mathrm{t'})\mathrm{dt'}=0
    \end{aligned}
\end{equation}
Now, using Coulomb's Gauge ($\nabla\cdot\mathbf{A}=0$), and using the ansatz, $\mathbf{A}(\mathbf{r},\mathrm{t})=\mathbf{A}_\mathrm{s}(\mathbf{r})+\mathbf{A}_\mathrm{t}(\mathbf{r},\mathrm{t})+\mathrm{c.c.}$, where $\mathbf{A}_\mathrm{t}(\mathbf{r},\mathrm{t})=\mathbf{A}_\mathrm{t0}(\mathbf{r})\mathrm{e}^{-\mathrm{i}\omega_\mathrm{f}\mathrm{t}}$, for the magnetic vector potential at a time after the time interface yields
\begin{equation}
    \begin{aligned}
        (\omega_\mathrm{p}^2-\varepsilon_\infty\omega_\mathrm{f}^2-\mathrm{c}^2\nabla^2)\mathbf{A}_\mathrm{t}+(\omega_\mathrm{p}^2-\mathrm{c}^2\nabla^2)\mathbf{A}_\mathrm{s}=\frac{\mathrm{e}^2}{\epsilon_0\mathrm{m_e}}\int_{-\infty}^\mathrm{t}\frac{\partial\mathrm{N^{(2)}_e}}{\partial \mathrm{t'}}\mathbf{A}(\mathrm{t'})\mathrm{dt'}
        \label{eq:Ampere}
    \end{aligned}
\end{equation}
Comparing the time-independent components on the right side of eq.~(\ref{eq:Ampere}) with the left side, we can derive a general differential equation to determine the magnitude of the static magnetic vector potential ($\mathbf{A}_\mathrm{s}$) and consequently the static magnetic field ($\mathbf{H}_\mathrm{s}$) at the hot spot corresponding to the QS mode as
\begin{equation}
\begin{gathered}
    (\omega_\mathrm{p}^2-\mathrm{c^2}\nabla^2)\mathbf{A}_\mathrm{s}(\mathbf{r})=\frac{\mathrm{e^2}}{\epsilon_0\mathrm{m_e}}\int_{-\infty}^\mathrm{t}\frac{\partial\mathrm{N^{(2)}_e}}{\partial\mathrm{t'}}\mathbf{A}(\mathbf{r},\mathrm{t'})\mathrm{dt'}\\\mathbf{H}_\mathrm{s}(\mathbf{r})=\frac{1}{\mu_0}\nabla\times\mathbf{A}_\mathrm{s}(\mathbf{r})
    \label{eq:A_s}
    \end{gathered}
\end{equation}
Furthermore, we can write the dispersion relation for propagating electromagnetic waves in the hot spot from eq.~(\ref{eq:Ampere}) as
\begin{equation}
    \nabla^2\mathbf{A}_\mathrm{t}=\frac{1}{\mathrm{c}^2}(\omega_\mathrm{p}^2-\varepsilon_\infty\omega_\mathrm{f}^2)\mathbf{A}_\mathrm{t}
    \label{eq:D_At}
\end{equation}
For a homogeneous plasma medium, the dispersion relation in eq.~(\ref{eq:D_At}) for a monochromatic wave propagating with a wave vector $\mathbf{k}_0$ reduces to the usual dispersion relation of an EM wave in a plasma and gives,
\begin{equation}
\varepsilon_\infty\omega_\mathrm{f}^2=\omega_\mathrm{p}^2+\mathrm{c^2}\mathbf{k}_0^2
\end{equation}

\bibliographystyle{ieeetr}
\bibliography{supp}